\documentclass{iopjournal}
\makeatletter
\renewcommand{\articletype}[1]{}
\makeatother

\usepackage{graphicx} 
\usepackage[ansinew]{inputenc}
\usepackage[bbgreekl]{mathbbol}
\usepackage{amsmath,amssymb,amsthm}
\usepackage{amsfonts}
\usepackage{bm}
\usepackage{bbm}
\usepackage{xcolor}

\usepackage{subcaption}  
\usepackage{cancel}

\newcommand{\fracp}[2]{\frac{\partial #1}{\partial #2}}
\newcommand{\fract}[2]{\frac{\mathrm{d} #1}{\mathrm{d} #2}}
\newcommand{\dd}{\mathop{}\!\mathrm{d}} 

\newcommand{\curl}{\nabla\times}

\newcommand\bk{\boldsymbol{k}}
\newcommand\bX{\boldsymbol{X}}

\newcommand\bx{\boldsymbol{x}}
\newcommand\bv{\boldsymbol{v}}
\newcommand\bE{\boldsymbol{E}}
\newcommand\bB{\boldsymbol{B}}
\newcommand\bA{\boldsymbol{A}}
\newcommand\bJ{\boldsymbol{J}}
\newcommand\bD{\boldsymbol{D}}
\newcommand\bH{\boldsymbol{H}}
\newcommand\bP{\boldsymbol{P}}

\newcommand{\Xd}{\dot{\bX}}
\newcommand{\vpar}{{v_\shortparallel}}

\newcommand{\Aex}{\boldsymbol{A}_{\text{eq}}}

\newcommand{\bex}{\boldsymbol{b}_{\text{eq}}}

\newcommand{\Bex}{\boldsymbol{B}_{\text{eq}}}
\newcommand{\Bzex}{B_{\text{eq}}}

\newcommand{\Btot}{\boldsymbol{B}_{\text{tot}}}

\newcommand{\Bpartot}{B_{\shortparallel,\text{tot}}}

\newcommand{\Bexm}{\left|\Bex \right|}

\newcommand{\Eperp}{\ensuremath{\bE_\perp}}

\newcommand{\Epar}{E_\shortparallel}

\newcommand{\Jepar}{J_{gc,\shortparallel}}

\newcommand{\Astar}{\boldsymbol{A}^\ast}
\newcommand{\Bstar}{\boldsymbol{B}^\ast}
\newcommand{\boldsymboltar}{\boldsymbol{B}^\ast}

\newcommand{\Bpstar}{{B_\shortparallel^\ast}}

\newcommand{\vgc}{\boldsymbol{v}_{gc}}
\newcommand{\agc}{a_{gc}}
\newcommand{\jgc}{\boldsymbol{J}_{gc}}

\newcommand{\pgcperp}{p_{\perp}}
\newcommand{\pgcpar}{p_{\shortparallel}}
\newcommand{\rhogc}{\rho_{gc}}

\newcommand{\kpar}{k_\shortparallel}
\newcommand{\kperp}{k_\perp}

\newtheorem{remark}{Remark}

\begin{document}
\articletype{Paper}

\title{A quasi-neutral electromagnetic hybrid model with drift-kinetic electrons and fully kinetic ions}

\author{Guo Meng$^{1,*}$, Nishant Narechania$^1$, Eric Sonnendr\"ucker$^{1,2}$}

\affil{$^1$ NMPP, Max Planck Institute for Plasma Physics, Garching, Germany}

\affil{$^2$ School of Computation Information and Technology, Technical University of Munich, Garching, Germany}

\email{guo.meng@ipp.mpg.de}


\begin{abstract}
In this work, we propose a hybrid model that combines drift-kinetic electrons with fully kinetic ions under the quasi-neutrality assumption, discretized using a geometric particle-in-cell framework on dual-grids. The model advances the perturbed electromagnetic fields $\bE$ and $\bB$ directly, rather than the scalar and vector potentials.  The parallel electric field $E_\parallel$ is obtained from Ohm's law. The perpendicular electric field $\bE_\perp$ is computed from Amp\`ere's law by extracting the $\bE_\perp$-dependent component of the drift-kinetic electron current. The quasi-neutrality constraint eliminates high-frequency light waves and Langmuir waves from the system.   
Temporal discretization is performed using low-storage Runge--Kutta schemes. In this quasi-neutral hybrid model, the right-hand polarized wave branch exhibits a whistler-like dispersion relation, which imposes a stringent timestep constraint.   To address this, we develop a novel implicit-explicit  splitting scheme for Faraday's law that significantly relaxes the timestep stability restriction. The model is validated in slab geometry by reproducing cold plasma wave branches, ion Bernstein waves, compressional and shear Alfv\'en waves, and ion acoustic waves.
\end{abstract}

\section{Introduction}

Kinetic descriptions of magnetized plasmas are essential for capturing multiscale phenomena such as wave-particle interactions, microinstabilities, and turbulence \cite{brizard2007foundations,chen2016RMP,garbet2010gkturb}. Fully kinetic Vlasov--Maxwell models provide the most comprehensive framework, but their computational cost is often prohibitive due to the disparate spatiotemporal scales between electrons and ions. To overcome this, hybrid kinetic approaches have been widely developed, with a notable class of models combining fully kinetic ions with reduced electron descriptions. In particular, gyro-kinetic (and its zero-Larmor-radius limit, drift-kinetic) electron models eliminate fast electron gyro-motion while preserving parallel dynamics and key kinetic effects such as Landau damping \cite{lin2005gefi}. Early formulations of such hybrid kinetic model can be found in \cite{YChen2009VlasovDrift}, where a particle-in-cell (PIC) framework coupling Vlasov ions with drift-kinetic electrons under quasi-neutral assumption was developed and applied to Alfv\'en wave physics. Extensions to electromagnetic regimes that retain displacement current and electron polarization effects have been proposed in \cite{chen2019gefiEB}, enabling the recovery of space-charge wave physics such as the lower-hybrid wave (LHW).
In addition to electromagnetic formulations, electrostatic hybrid models have also been extensively studied in fusion plasmas \cite{kuley2013verification,bao2014LHW}. For instance, a hybrid model with drift-kinetic electrons and fully kinetic ions, which uses a Poisson equation for the electrostatic potential, has been successfully applied to LHW physics in tokamaks \cite{bao2014LHW}.

More recently, structure-preserving discretizations have been introduced to improve the long-time fidelity of kinetic simulations following a discrete de Rham sequence and a particle description of the Vlasov equation \cite{kraus2017gempic,Kormann2024A-Dual-Grid, campos2022variational}. A geometric discretization based on Mimetic Finite Differences was extended to the drift-kinetic model and to a hybrid model with drift-kinetic electrons and fully kinetic ions \cite{Meng2025}. In this framework, the coupling to electromagnetic fields is formulated through a macroscopic Maxwell system incorporating polarization and magnetization effects induced by drift-kinetic particles. Furthermore, a geometrically discretized quasi-neutral Vlasov-Maxwell PIC model with fully kinetic electrons and ions was proposed in \cite{nishant2026}, where a Lagrange multiplier is introduced to enforce the discrete divergence of the current density to zero at machine precision.

An important challenge in electromagnetic kinetic simulations is the so-called ``cancellation problem'' arising in formulations based on scalar and vector potentials, particularly in simulations of shear Alfv\'en waves \cite{bao2018conservative,lu2025generalized}. This issue motivates the development of gauge-free formulations that evolve electromagnetic fields directly.
In this work, we develop the DeFi-QN (Drift-kinetic electrons and Fully kinetic ions, Quasi-Neutral) model by incorporating the quasi-neutrality assumption into the gauge-free hybrid kinetic framework \cite{Meng2025}. The proposed model is derived consistently from the drift-kinetic Vlasov equation, evaluating macroscopic moments directly from the distribution function rather than introducing an electron flow velocity. By neglecting the displacement current, the model explicitly filters out high-frequency light and Langmuir waves. Consequently, unlike formulations designed to capture LHW physics, our model operates in lower-frequency regimes.
The formulation evolves the electromagnetic fields $\mathbf{E}$ and $\mathbf{B}$ directly, avoiding the use of potentials and mitigating the ``cancellation problem'' associated with shear Alfv\'en waves (SAW). Furthermore, we retain the parallel magnetic perturbation $B_\parallel$, enabling a self-consistent description of compressional Alfv\'en waves (CAW). Together with the fully kinetic ion treatment, this framework enables studies of electromagnetic ion cyclotron range of frequencies (ICRF) waves and electrostatic ion Bernstein modes, including their nonlinear coupling.


The paper is organized as follows. Section~\ref{sec:DK_EB} presents the gauge-free drift-kinetic Vlasov-Maxwell model. In Section~\ref{sec:DeFi-QN}, this model is coupled with fully kinetic ions to derive the DeFi-QN field equations,  and a novel implicit-explicit (IMEX) splitting scheme for advancing $\bB$ is proposed. Section~\ref{sec:CPDR} derives the cold plasma dispersion relation and discusses the regime of validity of the model. Section~\ref{sec:benchmark} validates the model using the \texttt{GEMPICX} code \cite{gempicx} through comparisons with theoretical dispersion relations. Finally, Section~\ref{sec:conclusion} summarizes the main results and outlines future directions.

\section{The field based drift-kinetic model}\label{sec:DK_EB}

Consider a magnetized plasma with an equilibrium time-independent magnetic field $\Bex$.  Let $\bex$ denote the unit vector along the equilibrium magnetic field, $\bB$ the self-consistent magnetic field and $\bA$ the associated vector potential. The following modified vector potential and magnetic field are introduced
\begin{equation}\label{eq:Astar}
	\Astar= \bA + \Aex +\frac mq \vpar \bex,~~ \Bstar= \nabla\times\Astar = \bB +\Bex + \frac mq \vpar \nabla\times\bex, 
	~~ \Bpstar = \Bstar\cdot \bex.
\end{equation}
The modified magnetic field satisfies $\nabla \cdot \Bstar=0$,  and the phase-space Jacobian $\Bpstar$ satisfies a Liouville-type continuity equation in the characteristic phase space \cite{Meng2025}. 

We start from the drift-kinetic model described in \cite{Meng2025}, which consists of the drift-kinetic Vlasov equation,
\begin{equation}\label{eq:VlasovDK}
    \fracp{f}{t} + \fract{\bX}{t}\cdot\nabla_x f +  \fract{\vpar}{t}\fracp{f}{\vpar}=0,
\end{equation} 
together with the gyrocenter equations of motion
\begin{align}
	\fract{\bX}{t}&=\vpar\frac{\boldsymboltar}{\Bpstar} 
	+ \frac{1}{\Bpstar} \left(\bE\times\bex -\frac {\mu}{q} \nabla \Bpartot\times\bex \right) =: \vgc ,  \label{eq:Xdot}\\
	\fract{\vpar}{t}&= \frac {q}{m}\frac{\boldsymboltar}{\Bpstar} \cdot \left( \bE - \frac {\mu}{q} \nabla \Bpartot\right)  =: \agc .\label{eq:Vdot}
\end{align}
where $\bX$ denotes the gyrocenter position and $\vpar$ the parallel velocity. And $\vgc$ and $\agc$ are the gyrocenter velocity and acceleration, respectively, with $\vgc\cdot\bex=\vpar$. Here $\Bpartot=B_{\text{eq}}+\bB\cdot\bex$, where $B_{\text{eq}}=\Bexm$ is the magnitude of the equilibrium magnetic field. The total magnetic field is given by $\Btot=\Bex+\bB$. Consistent with the drift-kinetic ordering, the magnetic perturbation is small compared to the equilibrium field, i.e., $|\bB| \ll B_{\text{eq}}$. The particle dynamics are self-consistently coupled to the gyrokinetic Maxwell equations,
\begin{align}
    \fracp{\bD}{t} - \nabla \times \bH &=- \jgc , \label{eq:ampere}\\
    \fracp{\bB}{t} + \nabla \times \bE &= 0,\label{eq:faraday}\\
    \nabla \cdot \bD &= \rhogc , \label{eq:Gauss}\\
    \nabla \cdot  \bB &= 0,
\end{align}
with the electric displacement field $\bD$ and the magnetic field intensity $\bH$ defined as
\begin{align}
	\bD &= \epsilon_0\bE+\mathbf{P}, \text{ in the quasi-neutral limit},  \bD= \mathbf{P}  \label{def:D}\\
	\bH &= \frac{1}{\mu_0}(\Bex+\bB) - \mathbf{M}, \label{def:H}
\end{align}
where $\mathbf{P}$ and $\mathbf{M}$ are, respectively, the polarization and magnetization fields coming from the gyrokinetic approximation.
Specifically, the total polarization and magnetization  are the sum over the contributions from each species `$s$' treated gyrokinetically, $\mathbf{P}=\sum_s^{N_{DK}}\mathbf{P}_s$ and $\mathbf{M}=\sum_s^{N_{DK}}\mathbf{M}_s$. 
A local Maxwellian distribution function for drift-kinetic particles is given by
\begin{equation}\label{eq:Maxwellian}
	f_{M}(\bx,v_\shortparallel,\mu)=\frac{2\pi}{m}\frac{n_{M}(\bx)}{(2\pi)^{3/2} v_{th}^3 }\exp\left[-\frac{(v_\shortparallel-u_{\shortparallel}(\bx))^2}{2v_{th}(\bx)^2}-\frac{\mu|\Bex(\bx)|}{m v_{th}(\bx)^2}\right],
\end{equation}
where density $n_{M}(\bx)$, parallel mean flow velocity $u_{\shortparallel}$ and thermal velocity $v_{th}=\sqrt{  T/m}$, with $T$ the temperature of the species. Here, the distribution function is defined in the reduced $(v_\parallel,\mu)$ phase space, with the gyroangle integrated out, the factor $(2\pi/m)$ arising from the velocity-space Jacobian is absorbed into $f_M$. The normalization is thus $\int f_{M} B_{\text{eq}} \dd \vpar \dd \mu = n_{M}$.

Assuming a centered Maxwellian, polarization and magnetization are defined as
\begin{align}
	\mathbf{P}_s (t,\bx) &= \frac{m_s \, n_M}{B_{\text{eq}}^2}\bE_\perp =\frac{1}{\mu_0V_{A,s}^2}\bE_\perp, \label{eq:Plin}\\ 
	\mathbf{M}_s (t,\bx) &=   - \int \mu\bex f_s \Bpstar  \dd \vpar \dd \mu
    = - \frac{\pgcperp}{B_{\text{eq}}^2}\Bex = - \frac{\beta_\perp}{2\mu_0}\Bex,
    \label{eq:Mlin}  
\end{align}
where the Alfv\'en velocity is defined by
\begin{equation}
    V_{A,s}(\bx) = \frac{B_{\text{eq}}}{\sqrt{\mu_0 m_s n_{M}}},  \qquad \frac{1}{V_A^2}=\sum_s^{N_{DK}} \frac{1}{V_{A,s}^2},
\end{equation}
and the parallel pressure, perpendicular pressure, and plasma $\beta$ by
\begin{equation}
    \pgcpar = \int \vpar^2  f \Bpstar \dd \vpar\dd\mu,~~~~~
    \pgcperp = \int \mu \Bzex  f \Bpstar \dd \vpar\dd\mu, 
    ~~~~~ \beta = \frac{2\mu_0 p}{B_{\text{eq}}^2} .
\end{equation}
Note that we chose to keep the perturbed perpendicular pressure in the magnetization.

We define the gyro center charge density $\rhogc$ by 
\begin{equation}\label{eq:rhogc}
 \rhogc  = q \int   f \Bpstar   \dd \vpar \dd \mu,
\end{equation}
the gyro center current density by
\begin{equation}\label{eq:jgy}
\jgc = q \int  \vgc  f \Bpstar \dd \vpar \dd \mu.
\end{equation}
Finally, integrating the drift-kinetic Vlasov equation \eqref{eq:VlasovDK} over $\vpar$ yields the continuity equation
\begin{equation}
    \fracp{\rhogc}{t} + \nabla\cdot\jgc = 0.
\end{equation}

\section{Drift kinetic electrons coupling with fully-kinetic ions} \label{sec:DeFi-QN}
The field based drift-kinetic model naturally leads to a self-consistent coupling between the drift-kinetic and fully kinetic models within a hybrid framework. In the hybrid drift-kinetic electrons fully-kinetic ions model (DeFi), we get the following macroscopic Maxwell equations \cite{Meng2025},
\begin{align}
\fracp{\bD}{t} - \nabla \times \bH &= -(\jgc +\bJ_i) , \label{eq:ampereDKFK}\\
\fracp{\bB}{t} + \nabla \times \bE  &= 0,\label{eq:faradayDKFK}\\
\nabla \cdot \bD &= \rhogc + \rho_i , \label{eq:GaussDKFK}\\
\nabla \cdot  \bB &= 0.
\end{align}
Equations \eqref{def:D}--\eqref{def:H} with electrons treated as the only drift-kinetic species relate $\bD$ to $\bE$ and $\bH$ to $\bB$.

The fully kinetic ions obey the Lorentz equations  
\begin{align}
	\fract{\bx}{t}&=\bv, \label{eq:Xdoti}\\
	\fract{\bv}{t}&= \frac {q_i}{m_i}\left( \bE + \bv\times\Btot\right).\label{eq:Vdoti}
\end{align} 
We assume one species of ions with distribution function $f_i$ solving the fully kinetic Vlasov equation
\begin{equation}
    \fracp{f_i}{t} + \bv\cdot\nabla_x f_i +  \frac{q_i}{m_i}(\bE+\bv\times\Btot)\cdot\fracp{f_i}{\bv}=0.
\end{equation}
In general, one can include multiple particle species and choose either a drift-kinetic or a fully kinetic description for each of them. In hybrid models, it is physically natural to use a drift-kinetic formulation for electrons while treating ions fully kinetically, or to additionally include fully kinetic energetic particles or charged impurity species. 
For simplicity and without loss of generality, we use $e$ (or ${gc}$) and $i$ to denote drift-kinetic electrons (gyrocenters) and fully kinetic ions, respectively. The conclusions remain valid for more general multi-species hybrid models.
\subsection{Amp\`ere's law for the hybrid model}
Substituting $\bD$ and $\bH$ into Amp\`ere's law yields
\begin{align} \label{eq:ampere_of_DM}
\fracp{(\epsilon_0\bE+\mathbf{P}_e )}{t} -\nabla \times [\frac{1}{\mu_0}(\Bex+\bB) - \mathbf{M}_e]=-(\bJ_i+\jgc).
\end{align}
To evaluate the contribution of the electron magnetization, we define the electron magnetization current density as $\mathbf{J}_{M_e} = \nabla \times \mathbf{M}_e$. For a Maxwellian distribution, the electron pressure is isotropic: $p_{e,\perp}=p_{e,\|}=p_e=n_eT_e$.
The magnetization current therefore satisfies
\begin{equation}
    \mathbf{J}_{M_e} = -\nabla\times (\frac{p_{e,\perp}}{B_{\text{eq}}^2}\Bex ) =- \nabla (\frac{  p_{e,\perp}} {B_{\text{eq}}^2})\times \Bex-\frac{ p_{e,\perp}} {B_{\text{eq}}^2} \curl \Bex.
\end{equation}
Physically, $\mathbf{J}_{M_e}$ is predominantly perpendicular to the equilibrium magnetic field, contributing primarily to the diamagnetic response. Its parallel component is proportional to the magnetic twist $\bex \cdot (\nabla \times \bex)$, which vanishes identically in vacuum or current-free configurations where $\nabla \times \Bex = 0$. This vacuum-field approximation is valid for stellarator equilibria, it implies that $\mathbf{J}_{M_e}$ does not contribute to the parallel electric field $E_\parallel$ equation. In contrast, tokamak equilibria involve a poloidal field $B_\theta$ generated by the plasma current, results in a non-vanishing $\nabla \times \Bex$; however, $\bex\cdot \mathbf{J}_{M_e}$ remains small in the low-$\beta$ limit and is often neglected in studies of fast parallel dynamics \cite{brizard2007foundations}. The expansion of the first term on the right-hand side yields, 
\begin{equation}
    - \nabla (\frac{  p_{e,\perp}} {B_{\text{eq}}^2})\times \Bex = \frac{\Bex \times  \nabla p_{e,\perp}} {B_{\text{eq}}^2} + \frac{2 p_{e,\perp}}{B_{\text{eq}}^2} (\nabla B_{\text{eq}} \times \mathbf{\bex}) .
\end{equation}
where the first term is the electron diamagnetic current, $\bJ_{e,{\rm dia}}=\frac{\Bex \times \nabla p_e}{B_{\text{eq}}^2}$. The second term represents a correction arising from the magnetic field gradient. The component $\nabla\times(\frac{1}{\mu_0}\Bex -\mathbf{M}_e)$ is critical for model consistency. Specifically, since the electron guiding-center velocity $\vgc$ does not include the diamagnetic drift, the magnetization term must account for this. The current density arising from electron grad-B drift motion in the equilibrium field is
$$
    \mathbf{J}_{gc, \nabla B} = n_e q_e\langle \mathbf{v}_{gc,\nabla B} \rangle = n_e \left( \frac{\mu}{\Bzex} \bex \times \nabla \Bzex \right) = \frac{p_{e,\perp}}{\Bzex^2} \bex \times \nabla \Bzex.
$$ Summing this guiding-center contribution with the magnetization correction term, $\frac{2 p_{e}}{\Bzex^2} (\nabla \Bzex \times \bex)$, recovers the correct net fluid current related to the magnetic field gradient.
In contrast, the ion dynamics is described using a fully kinetic model. Here, the ion diamagnetic current is naturally contained in $\bJ_i$, since the velocity moment $\bJ_i=q_i\int \bv f_i dv^3 $ includes the diamagnetic flow, which emerges from the orbit-averaged effect of finite gyroradius sampling of a spatially nonuniform equilibrium distribution, $\bJ_{i,{\rm dia}} = \frac{\Bex \times \nabla p_i}{B_{\text{eq}}^2}$ \cite{littlejohn1983variational,brizard2007foundations}. Consequently, the equilibrium force balance is implicitly satisfied in Eq.~\eqref{eq:ampere_of_DM}, since the full-$f$ expression is employed. 
We therefore get 
\begin{align} \label{eq:ampere_Jedia}
\fracp{(\epsilon_0\bE+\mathbf{P}_e )}{t} -\frac{1}{\mu_0} \nabla \times \bB   =  \frac{1}{\mu_0} \nabla \times \Bex-(\bJ_i+\jgc + \nabla\times \mathbf{M}_e).
\end{align}
Note that this $\mathbf{P}_e$ is perpendicular and will henceforth not be in the $\Epar$ equation of this quasi-neutral model. This equation can be rewritten by separating the perpendicular and parallel electric-field components as

\begin{equation}\label{linearampere}
    \fracp{\bE }{t}  +\frac{c^2}{V_{A,e}^2}\fracp{\Eperp}{t}- c^2 \nabla \times \bB =-\frac{1}{\epsilon_0}(\bJ_i+\jgc)+c^2 \nabla \times \Bex-\frac{1}{\epsilon_0}\nabla\times \mathbf{M}_e.
\end{equation}
The first term represents the displacement current, while the second term corresponds to the electron polarization current. The ratio of their magnitudes is given by $\displaystyle \frac{c^2}{V_{A,e}^2}=\frac{\omega_{pe}^2}{\Omega_{ce}^2}$, where $\omega_{pe}=\sqrt{n_M q_e^2/(\epsilon_0m_e)}$ is the electron plasma frequency and $\Omega_{ce}=-e \Bzex/m_e$ is the electron cyclotron frequency. 
Because this ratio can be of order unity in typical high-density plasmas (e.g., fusion-relevant regimes where $c/V_{A,e}\sim 1$), these two terms must be treated on an equal footing. Consistent with standard gyrokinetic ordering, both terms are formally $\mathcal{O}(m_e/m_i)$ smaller than the dominant ion polarization current and are therefore simultaneously neglected to maintain theoretical self-consistency.

\subsection{The quasi-neutral model without electron polarization}
Quasi-neutrality implies that the total charge density in the plasma vanishes, \textit{i.e.} $\sum_s q_s n_s = 0$, where the sum is over all the particle species in the plasma. Via the continuity equation this implies that $\nabla\cdot\bJ=0$, where $\bJ$ is now the total current summed up over all species, $\bJ= \bJ_i + \jgc$. Following the scaling considerations in \cite{Degond2017AP-review}, we observe that the quasi-neutral approximation is obtained by taking the limit of Amp\`ere's law when $\epsilon_0$ goes to zero. In our model this consists in setting $\bD = \bP $. This implies in particular that the parallel component of $\bD$ vanishes, \textit{i.e.} $\bD\cdot\bex=0$. This changes the nature of the equations as the Amp\`ere equation does not involve $E_\parallel$ in the time derivative, but on the other hand taking its dot product with $\bex$ yields the following constraint on the magnetic field
\begin{equation}\label{eq:constraintAmpere}
    \bex\cdot \nabla\times\bH =  \bex \cdot \bJ. 
\end{equation} 

If the polarization term $\frac{1}{V_{A,e}^2}\fracp{\Eperp}{t}$ is kept in the Amp\`ere law, the quasi-neutral form of the Amp\`ere's equation including the electron polarization is
\begin{equation}\label{eq:ampereQN_hybrid}
   -\frac{1}{V_{A,e}^2}\fracp{\bE_\perp}{t} + \nabla\times\bB = \mu_0(\bJ_i + \jgc+\nabla\times \mathbf{M}_e)-\nabla\times \Bex,
\end{equation}
where $V_{A,e}$ is the electron Alfv\'en velocity. 
Taking the time derivative of the Amp\`ere's equation Eq.~\eqref{eq:ampereQN_hybrid} and substituting it into the curl of Faraday's equation Eq.~\eqref{eq:faraday}, we get 
\begin{equation} \label{eq:qncurlcurl_hybrid}
    \frac{1}{V_{A,e}^2}\frac{\partial^2\bE_\perp}{\partial t^2} + \nabla \times \nabla \times \bE = - \mu_0 \fracp{(\bJ_i + \jgc +\nabla\times \mathbf{M}_e)}{t}.
\end{equation}
Additionally, since $\Bex$ is defined as equilibrium, its time derivative vanishes. The DeFi model with polarization and displacement current has been discussed in Ref. \cite{chen2019gefiEB}, and it allows LHW to be retained in the dynamics. In this work, the electron polarization current is neglected. This approximation is well justified for a realistic ion-to-electron mass ratio, for which the electron polarization contribution is parametrically small in the low-frequency regime of interest. Then Amp\`ere's equation reduces to
\begin{equation}\label{eq:ampereQN}
    \nabla\times\bB = \mu_0(\bJ_i + \jgc+\nabla\times \mathbf{M}_e )-\nabla\times \Bex,
\end{equation}
along with the quasi-neutrality condition $$\rho=\rho_i+\rhogc=0,$$ which implies current continuity, $$\nabla\cdot(\bJ_i+\jgc)=0.$$
The quasi-neutral approximation and the neglect of the displacement current are adopted to consistently remove high-frequency electromagnetic modes from the system. When the electron polarization current is retained, high-frequency branches such as the L-mode and the lower-X mode appear, and their frequencies are not correctly captured even in the long-wavelength (low-$k$) limit. Since these modes are not relevant to the low-frequency electromagnetic instabilities of interest here, they are intentionally excluded from the model. We will show this later in Section \ref{sec:CPDR}, where cold plasma dispersion relations are derived for different models.

The quasi-neutral Amp\`ere equation Eq.~\eqref{eq:ampereQN} together with the Faraday equation Eq.~\eqref{eq:faraday}, constitute the fundamental equations of the DeFi-QN (drift-kinetic electron and fully-kinetic ion with quasi-neutrality) model.

\subsection{Equations for $\Epar$}
By applying the dot product $\bex \cdot $ to equation \eqref{eq:qncurlcurl_hybrid}, we obtain
\begin{equation}
   \bex \cdot \nabla \times \nabla \times \bE = - \mu_0 \fracp{(\bJ_i + \jgc)}{t} \cdot \bex. 
\end{equation}
While the magnetization $\mathbf{M}_e$ defined in Eq.~\eqref{eq:Mlin} depends on the full distribution function $f_e$, the term  $\partial (\bex \cdot \nabla \times \mathbf{M}_e)/\partial t= -\frac{\partial p_{e,\perp}}{\partial t} \frac{\bex \cdot  \curl \Bex} {B_{\text{eq}}^2} $is neglected here for two reasons. First, the equilibrium pressure profile evolves on a slow transport timescale compared to the fast plasma dynamics of interest. Second, in the drift-kinetic limit, higher-order dynamic contributions from the perturbation $\delta f_e$ to the magnetization current are typically $O(\rho_e/L)$ smaller than the guiding-center current response and are thus negligible for fast plasma dynamics. 
From the fully kinetic Vlasov equation for ions, we derive 
\begin{equation}\label{eq:FKdjdt}
    \fracp{\bJ_i}{t} = \sum_i \frac{q_i}{m_i} (\rho_i\bE + \bJ_i \times \Btot - \nabla \cdot \mathbb{S}_i),
\end{equation}
where the charge density, current and stress tensor are given by \cite{nishant2026}
\begin{gather}
    \rho_i =  q_i \int f_i d\bv, \label{eq:rho_i}\\
    \bJ_i = q_i \int f_i \bv d\bv, \label{eq:J_i}\\
    \mathbb{S}_i = m_i \int f_i \bv \otimes \bv d\bv. \label{eq:S_i}
\end{gather}
For the drift-kinetic electrons, we denote the parallel component of $\jgc$ by $\Jepar$,
\begin{equation}\label{eq:jpar}
\Jepar = q_e \int  \vpar  f_e \Bpstar \dd \vpar \dd \mu.
\end{equation} 
To derive $\fracp{\Jepar}{t}$, we use the drift-kinetic Vlasov equation in conservative form \cite{Meng2025}
\begin{equation}
	\fracp{f_e\Bpstar}{t} + \nabla\cdot (f_e\Bpstar \Xd ) + \fracp{(f_e\Bpstar \dot{\vpar}) }{\vpar} =0,
\end{equation}
which can be written explicitly as
\begin{equation}\label{eq:VlasovDHCons}
   \fracp{f_e \Bpstar}{t} + \nabla\cdot  f_e\left( \vpar\Bstar + \Eperp\times\bex - \frac{\mu}{q_e} \nabla \Bpartot\times\bex\right) + \fracp{}{\vpar}f_e\left( \frac{q_e}{m_e} \Bstar\cdot\bE - \frac{\mu}{m_e}\Bstar\cdot\nabla\Bpartot\right) =0.
\end{equation}
Multiplying by $q_e \vpar$ and integrating over $\vpar$ and $\mu$, yields the following equation
\begin{equation}\label{eq:djpardt}
    \fracp{\Jepar}{t} + \nabla\cdot \left[\boldsymbol{p}^\ast + (\eta\bE- Q\nabla \Bpartot)\times\bex\right]
    - \boldsymbol{\alpha}^\ast \cdot \bE +\boldsymbol{\mu}^\ast\cdot\nabla\Bpartot
    = 0
\end{equation}
We define the following moments:
\begin{align}
	\boldsymbol{p}^\ast(\bx) &=   q_e \int \vpar^2 f_e \Bstar \dd \vpar\dd\mu, \\
	\eta(\bx) &=   q_e \int \vpar  f_e  \dd \vpar\dd\mu , \\
    Q(\bx) &=  \int \vpar\mu  f_e  \dd \vpar\dd\mu, \\
    \boldsymbol{\alpha}^\ast(\bx) &=   \frac{q_e^2}{m_e} \int f_e\Bstar \dd \vpar\dd\mu, \\
    \boldsymbol{\mu}^\ast(\bx) &=   \frac{q_e}{m_e} \int \mu f_e \Bstar \dd \vpar\dd\mu.
\end{align}
The divergence term satisfies
\[
\nabla \cdot \left[ \boldsymbol{p}^\ast + (\eta \bE - Q \nabla \Bpartot)\times\bex \right]
= \frac{q_e}{m_e} (\nabla \cdot \mathbb{S}_e)\cdot \bex,
\]
where the electron stress tensor is given by
\begin{equation} \label{eq:S_e}
\mathbb{S}_e = m_e \int f_e \, \bv_{gc} \otimes \bv_{gc} \, \Bpstar \, d\vpar \, d\mu.
\end{equation}
The contribution $(\nabla \cdot \mathbb{S}_e)\cdot \bex$ is primarily determined by $\bex \cdot \nabla p_{e \parallel}$, as the gyrocenter velocity $\bv_{gc}$ is dominated by its parallel component $\vpar$. Note the $\Epar$ term in $\boldsymbol{\alpha}^\ast\cdot \bE$
$$\boldsymbol{\alpha}^\ast\cdot\bex =  \frac{q_e^2}{m_e} \int f_e\Bpstar \dd \vpar\dd\mu 
= \frac{q_e^2 n_e(\bx)}{m_e} =: \alpha_e(\bx)=\epsilon_0 \omega_{pe}^2.$$ Combining the above results, we obtain the equation for $\Epar$:
\begin{multline}
\label{eq:curlcurlE}
    \bex \cdot \nabla \times \nabla \times \bE + \mu_0 (\sum_i \frac{q_i^2}{m_i}n_i + \frac{q_e^2}{m_e}n_e ) \Epar = \mu_0 \sum_i \frac{q_i}{m_i} (-\bJ_i \times \bB + \nabla \cdot \mathbb{S}_i) \cdot \bex \\
    + \mu_0\left\{ \nabla\cdot \left[\boldsymbol{p}^\ast + (\eta\Eperp- Q\nabla \Bpartot)\times\bex\right]
    - \boldsymbol{\alpha}^\ast \cdot \Eperp +\boldsymbol{\mu}^\ast\cdot\nabla\Bpartot\right\}
\end{multline}
Here we used the identity $(\bJ_i \times \Btot)\cdot\bex= (\bJ_i \times \bB)\cdot\bex$.  Note that the $\nabla \times \nabla \times \bE$ term contains a contribution from $\Eperp$. Assuming $\bB$, $\Eperp$ and $f_e$, $f_i$,  or equivalently the particle positions are known, $\Epar$ can be determined. For a constant unit vector $\bex$, we have $\bex\cdot \nabla\times\nabla\times\bE = -\nabla_\perp^2 E_\parallel + \bex \cdot (\nabla(\nabla\cdot\Eperp)).$

\subsection{Equations for \texorpdfstring{\Eperp}{Eperp}}
The $\Eperp$ equation is derived directly from the quasi-neutral Amp\`ere Eq. \eqref{eq:ampereQN}. As we noticed the $\Eperp$ is in the gyrocenter velocity equation Eq.~\eqref{eq:Xdot}.
Substituting $\vgc$ then we get the explicit $\jgc$ of electrons 
\begin{equation}\label{eq:jgc}
    \jgc =q_e\int \vpar\Bstar f_e \dd\vpar\dd\mu + \int (q_e\Eperp - \mu\nabla\Bpartot)\times\bex f_e  \dd\vpar\dd\mu. 
\end{equation}
As $\Eperp$ is contained in the expression of $\jgc$ from \eqref{eq:jgc}, substituting Eq. \eqref{eq:jgc} into \eqref{eq:ampereQN_hybrid}, we get the following evolution equation for $\Eperp$
\begin{multline}
        -\frac{1}{V_{A,e}^2}\fracp{\bE_\perp}{t} + \nabla\times (\bB+\Bex) = \mu_0(\bJ_i+\curl \mathbf{M}_e) + \mu_0 q_e\int \vpar\Bstar f \dd\vpar\dd\mu \\ + \mu_0\int (q_e\Eperp - \mu\nabla\Bpartot)\times\bex f  \dd\vpar\dd\mu. 
\end{multline}
When the electron polarization current is neglected, the first term vanishes and we can
derive an equation for solving $\Eperp$
\begin{multline} \label{eq:Eperp_1}
 \mu_0 q_e \int  \Eperp \times\bex f_e  \dd\vpar\dd\mu   = \nabla\times(\bB +\Bex)- \mu_0(\bJ_i+\curl \mathbf{M}_e ) \\- \mu_0  q_e \int \left( \vpar\Bstar - \frac{\mu}{q_e}\nabla\Bpartot \times\bex \right)  f_e \dd\vpar\dd\mu . 
\end{multline}
Using $\bex \times (\Eperp \times \bex)= (\bex \cdot \bex) \Eperp -(\bex \cdot \Eperp) \bex = \Eperp$, we get
\begin{equation} \label{eq:Eperp_2}
\begin{aligned}
 \mu_0 q_e \int \frac{1}{\Bpstar} \Eperp  f \Bpstar \dd\vpar\dd\mu   & = \bex \times \left[\nabla\times(\bB+\Bex) - \mu_0(\bJ_i+\curl \mathbf{M}_e) \right] \\
  & - \mu_0  q_e \int  \frac{1}{\Bpstar} \left( \vpar \bex \times \Bstar - \frac{\mu}{q_e}\nabla_\perp \Bpartot  \right)  f_e \Bpstar \dd\vpar\dd\mu, 
 \end{aligned}
\end{equation}
or equivalently
\begin{equation} \label{eq:Eperp}
 \mu_0  \varrho_e \Eperp  = \bex \times \left[\nabla\times (\bB+\Bex )- \mu_0(\bJ_i+\curl \mathbf{M}_e)\right] - \mu_0  q_e \int   \left( \vpar \bex \times \Bstar - \frac{\mu}{q_e}\nabla_\perp \Bpartot  \right)  f_e  \dd\vpar\dd\mu . 
\end{equation}
where we denote $\varrho_e = q_e\int f_e\dd\vpar\dd\mu$, which is not the same as ${\rhogc}_e$ as $\Bpstar$ is missing.
This is an explicit expression for $\Eperp$ assuming $\bB$, $f_e$, and $f_i$ or equivalently the particle positions are known.
For a uniform background field, the equilibrium pressure $p_{e,0}$ is constant. By substituting the magnetization current $\curl \mathbf{M}_e=\frac{\Bex \times \nabla p_{e,\perp}}{B_{\text{eq}}^2}$, the equation simplifies to
\begin{equation} \label{eq:Eperp_with_Me}
 \mu_0  \varrho_e \Eperp  = \bex \times (\nabla\times\bB - \mu_0\bJ_i) +\mu_0 \frac{\nabla_\perp \delta p_{e,\perp}}{\Bzex}  - \mu_0  q_e \int   \left( \vpar \bex \times \Bstar - \frac{\mu}{q_e}\nabla_\perp \Bpartot  \right)  f_e  \dd\vpar\dd\mu.
\end{equation}
Although formulated differently to suit our hybrid framework, this expression is physically consistent with the governing equations derived in Ref.~\cite{YChen2009VlasovDrift}.

In the following sections, for initial implementations in a uniform plasma, we restrict attention
to $\mu=0$ and thus the magnetization $ \mathbf{M}_e=0$.

\subsection{Implicit-explicit method for advancing \texorpdfstring{$\bB$}{B}}
We use the low-storage Runge--Kutta (LSRK) schemes for the time discretization and a dual-grid approach for the spatial discretization \cite{Meng2025}. At each substage of a time step, first $\Eperp$ is computed from \eqref{eq:Eperp}, and $\Epar$ is computed from \eqref{eq:curlcurlE}. The particles are then advanced, after which the $\bB$ is updated using Faraday's equation \eqref{eq:faraday}. To overcome the stringent time step constraint imposed by the fast wave branch, a splitting strategy based on an implicit-explicit (IMEX) scheme is employed for advancing the magnetic field $\bB$. By substituting the expression for $\Eperp$ from \eqref{eq:Eperp} into Faraday's law \eqref{eq:faraday}, an evolution equation involving only $\bB$ is obtained. The $\nabla\times \bB$ term in \eqref{eq:Eperp} will yield a stiff term in the equation for $\bB$, which can be treated implicitly, while the other terms can be treated explicitly. This can be done with a splitting scheme or an IMEX (implicit-explicit) Runge-Kutta method, where the stiff term is treated implicitly, for example with the Crank-Nicolson scheme, and the other terms explicitly using a LSRK method of order 3 or 4 to ensure stability for eigenvalues on the imaginary axis (oscillatory system $\lambda=i\omega$).

The equation treated implicitly is
\begin{equation}\label{eq:B_implicit}
    \fracp{\bB}{t} + \nabla\times\left(\frac{1}{\mu_0 \varrho_e} \bex \times ( \nabla\times\bB) \right)= 0.
\end{equation}
In an initial step, $\varrho_e$ may be approximated by the background quantities and the system which conserves the magnetic energy can be solved with a Crank-Nicolson scheme. Note that $\varrho_e = q_e\int f_e\dd\vpar\dd\mu$ is not equal to the gyrocenter charge density ${\rhogc}_e$ as $\Bpstar$ is missing. Therefore $\varrho_e \neq n_e q_e$, but instead scales as $\varrho_e  \sim n_e q_e/\Bzex$. The stiff term involves second-order derivatives in the parallel direction, leading to a restrictive time step condition. In particular, the Courant-Friedrichs-Lewy (CFL) constraint imposes  $\Delta t \leq C_{max} \frac{\mu_0 e n_e } {\Bzex}\Delta z^2$ which can be highly limiting in practice for explicit schemes. By treating this part implicitly, however, this severe time step restriction can be significantly relaxed, allowing for stable integration at larger time steps.

All the other terms in the Faraday equation can be treated explicitly, along with the particle pushing. These contributions satisfy
\begin{equation}\label{eq:B_explicit}
    \fracp{\bB}{t} - \nabla\times\left[ \frac{1}{\varrho_e} \bex \times \bJ_i +  \frac{q_e}{\varrho_e} \int   \left( \vpar \bex \times \Bstar - \frac{\mu}{q_e}\nabla_\perp \Bpartot  \right)  f  \dd\vpar\dd\mu \right] + \nabla\times(E_\parallel \bex)= 0.
\end{equation}
Equations~\eqref{eq:B_implicit} and \eqref{eq:B_explicit} together define the complete update of the magnetic field $\bB$. All terms appearing in Eq.~\eqref{eq:B_explicit} are already evaluated during the computation of Eqs.~\eqref{eq:curlcurlE} $\Epar$ and ~\eqref{eq:Eperp} $\Eperp$; therefore, no additional calculations are required. Equations~\eqref{eq:B_implicit}-\eqref{eq:B_explicit} merely represent a decomposition of Faraday's law into stiff and non-stiff components.

With the IMEX scheme, the time step restriction is significantly relaxed, following the scaling $\Delta t \propto  \Delta z$. A detailed description of geometric discretization, analysis of numerical stability and convergence will be presented in another work \cite{nishant_cpc}.
\begin{remark}
    Note that when keeping the electron polarization current, an Asymptotic-Preserving scheme for the full system can be designed by treating the stiff term coming from $\Eperp$ implicitly in Amp\`ere's equation. When $V_A\to \infty$, this will yield directly the quasi-neutral model without electron polarization current.
\end{remark}
\subsection{The stiff term and Hall effect}
The stiff term corresponds to the Hall effect. In the generalized Ohm's law, the classical Hall electric field is given by
\begin{equation}
    \bE_{\text Hall}= \frac{1}{n_e e} \bJ_{total} \times\Btot \approx \frac{1}{n_e e} \bJ_{total} \times\Bex,
\end{equation} 
since the  quasi-neutral Amp\`ere's law Eq. \eqref{eq:ampereQN} gives $\bJ_{total}=\frac{1}{\mu_0}\nabla\times \bB$. When the Hall electric field is inserted into Faraday's law, the stiff operator generates the characteristic whistler-wave (WHW) dynamics.  
Starting from the Hall contribution to the perpendicular electric field,
\[
    \Eperp^{\rm Hall} 
    = \frac{1}{\mu_0 \varrho_e}\,
      \bex \times (\nabla\times\bB),
    \qquad 
    \varrho_e = n_e q_e/\Bzex,
    \qquad q_e=-e.
\]
Insert this term into Faraday's law becomes
\[
    \frac{\partial \bB}{\partial t}
    = -\,\nabla\times\!\left(
        \frac{1}{\mu_0 \varrho_e}\,
        \bex\times(\nabla\times\bB)
      \right).
\]
In a uniform background field 
$\bex = \hat{\mathbf{z}}$, with constant density
$\varrho_e \sim n_e q_e/\Bzex$.  
Applying a Fourier ansatz 
$\bB \propto e^{i(\mathbf{k}\cdot\mathbf{x}-\omega t)}$,
the resulting dispersion relation is
\[
    \omega = \pm 
    \frac{\Bzex}{\mu_0 n_e e}\,
    k_\parallel\, k,
\]
where $k=\sqrt{k_\parallel^2+k_\perp^2}$. Here we have used $\nabla\cdot \bB=0 \Rightarrow \bk \cdot \hat{\bB}=0$. 
This wave is dispersive, as its phase velocity depends on $k$. For $\kperp=0$, $\omega \propto k^2$,
which is the well-known whistler scaling. 
Thus the stiff term is related to the Hall response of the electrons and produces the stiff
high-frequency, highly dispersive whistler waves.
The parameter $A=\frac{\Bzex}{\mu_0 n_e e}$ can be expressed in terms of standard electron plasma parameters as $A=\frac{V_{Ae}^2}{|\Omega_{ce}|}=|\Omega_{ce}|\frac{c^2}{\omega_{pe}^2}=|\Omega_{ce}|d_e^2$, where $d_e=c/\omega_{pe}$  is the electron inertial length (electron skin depth). The relationship between this stiff term and the associated whistler-wave dynamics is further illustrated in Figs.
\ref{fig:DeFi_QN_kpara} and \ref{fig:DeFi-QN_theta60}.

\section{Cold plasma dispersion relation and validation regime of the DeFi models} \label{sec:CPDR}
In this section, we benchmark the proposed formulation against the analytical dispersion relation of a uniform cold plasma. This choice allows the physical behavior of the model and its numerical stability properties to be examined in a controlled and transparent manner.

\begin{figure}[htbp]
    \centering
    \begin{subfigure}[b]{0.32\textwidth}
        \centering
        \includegraphics[width=\linewidth]{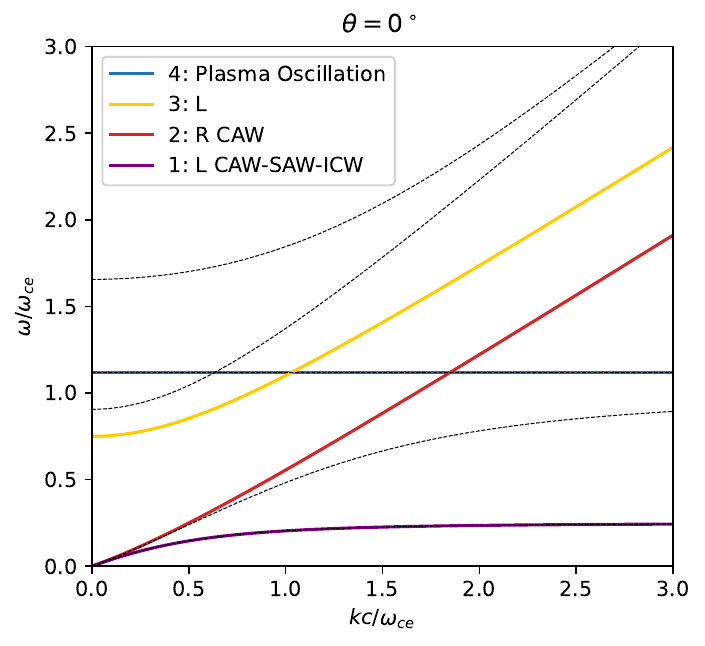}
        \caption{DeFi, $k_\parallel$.}
        \label{fig:DeFi_kpar}
    \end{subfigure}
    \hspace{2pt} 
    \begin{subfigure}[b]{0.32\textwidth}
        \centering
        \includegraphics[width=\linewidth]{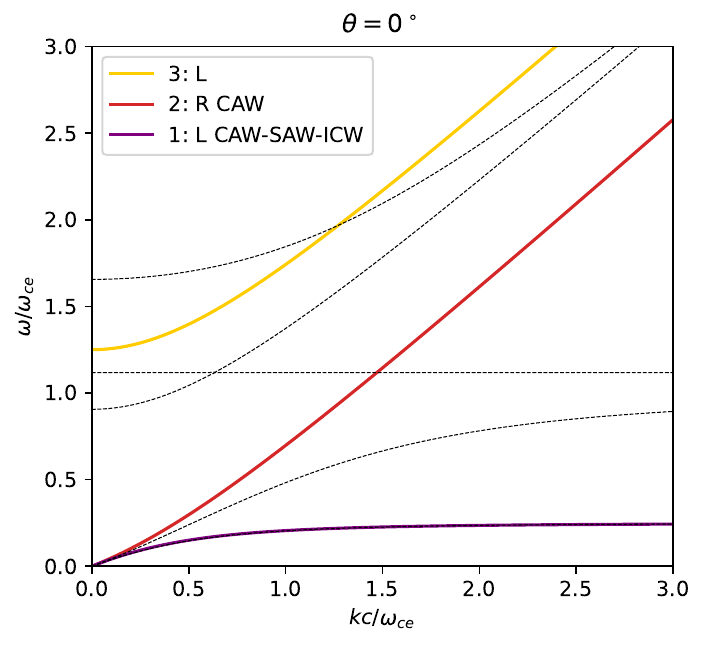}
        \caption{DeFi-QN-EP, $k_\parallel$.}
        \label{fig:DeFi-QN-EP_kpar}
    \end{subfigure}
    \hspace{2pt} 
    \begin{subfigure}[b]{0.32\textwidth}
        \centering
        \includegraphics[width=\linewidth]{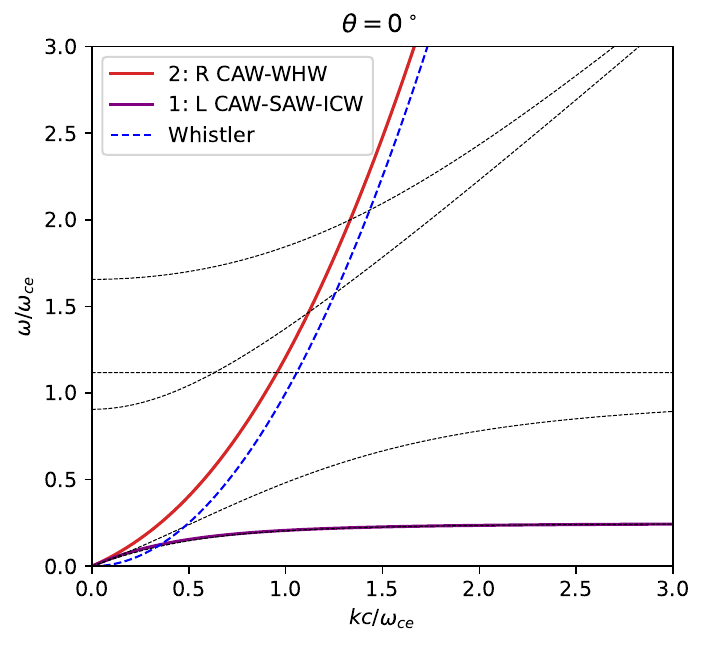}
        \caption{DeFi-QN, $k_\parallel$.}
        \label{fig:DeFi_QN_kpara}
    \end{subfigure}

    \begin{subfigure}[b]{0.32\textwidth}
        \centering
        \includegraphics[width=\linewidth]{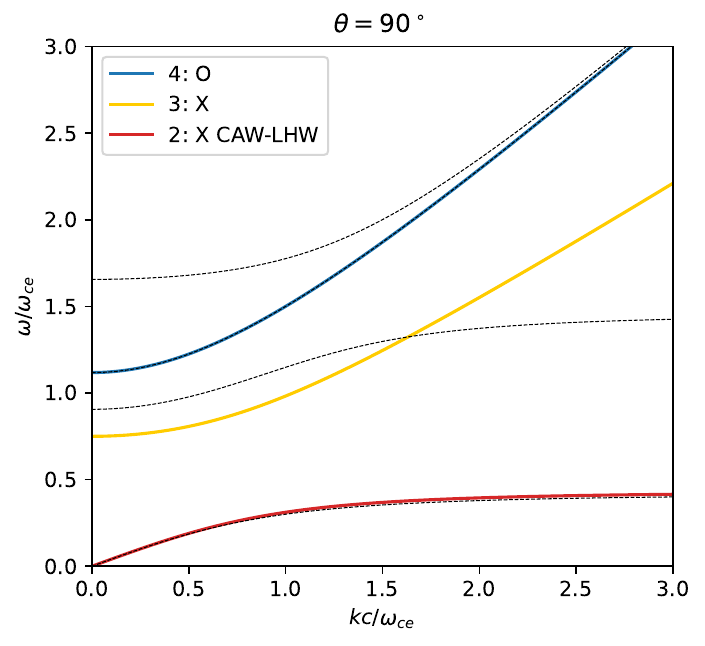}
        \caption{DeFi, $k_\perp$.}
        \label{fig:DeFi_kperp}
    \end{subfigure}
    \hspace{2pt} 
    \begin{subfigure}[b]{0.32\textwidth}
        \centering
        \includegraphics[width=\linewidth]{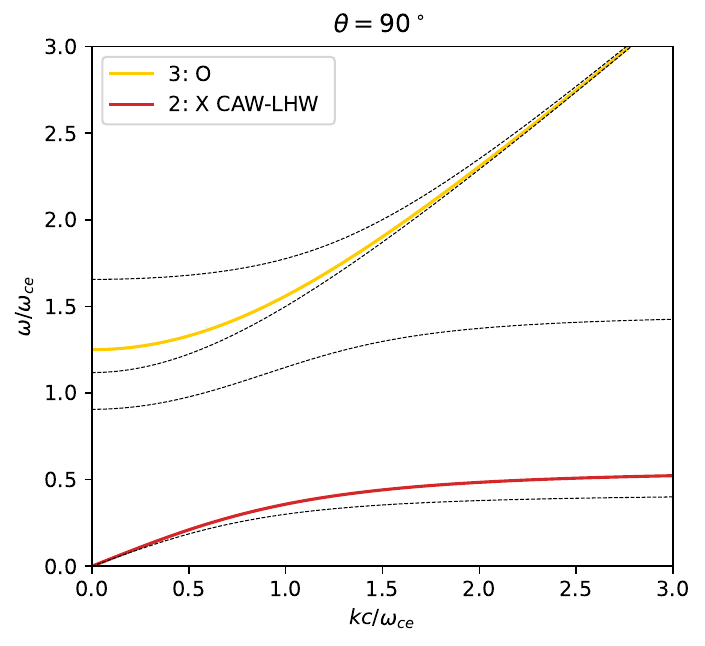}
        \caption{DeFi-QN-EP, $k_\perp$.}
        \label{fig:DeFi-QN-EP_kperp}
    \end{subfigure}
    \hspace{2pt} 
    \begin{subfigure}[b]{0.32\textwidth}
        \centering
        \includegraphics[width=\linewidth]{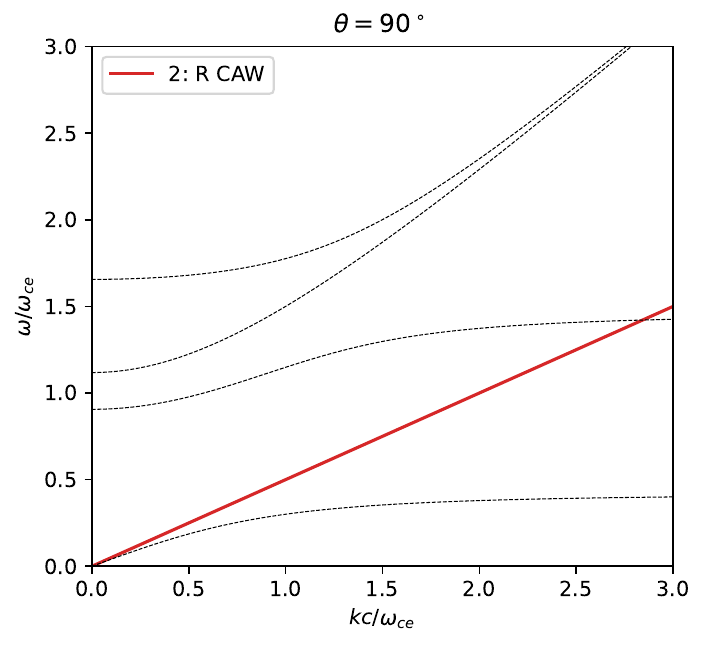}
        \caption{DeFi-QN, $k_\perp$.}
        \label{fig:DeFi_QN_kperp}
    \end{subfigure}
    \caption{Comparison of wave branches in different hybrid models for $m_i/m_e=4$,  $\omega_{pe}/|\Omega_{ce}|=1$ and $T_i=T_e$.  Left: DeFi model including both the displacement current and the electron polarization current.
  Middle: DeFi-QN model with electron polarization (EP) current.
  Right: DeFi-QN model without electron polarization current. 
  Black dashed lines indicate five ``full-CPDR" branches. Colored solid lines represent the eigenmode branches of different hybrid models. Their overlap demonstrates that the hybrid formulations capture the correct wave physics.   In the right upper panel, the blue dashed line shows the whistler-wave dispersion relation $\omega=|\Omega_{ce}|\frac{c^2}{\omega_{pe}^2} \kpar^2$.}
    \label{fig:compare_hybrid_CPDR}
\end{figure}

We consider a uniform plasma in a constant external magnetic field of the form $\Bex = (0,0,\Bzex)^\top$. The plasma consists of two species, electrons and ions (charge $q_i=e$), with equal densities \(n_i = n_e\), so that $|\Omega_{ce}/\Omega_{ci}|=m_i/m_e$.
The dispersion relation can be written in the compact form $\boldsymbol{\Lambda}(\boldsymbol{k},\omega)\bE= (1+\boldsymbol{\chi})\bE + \frac{c^2}{\omega^2} \boldsymbol{k}\times\boldsymbol{k}\times \bE=0$, where $\boldsymbol{\chi}$ is the susceptibility tensor.
For the cold plasma dispersion relation (CPDR), the dispersion tensor is commonly expressed using Stix's notation \cite{stix1992waves}
\begin{equation}
\boldsymbol{\Lambda}(\boldsymbol{k}, \omega) = 
\begin{bmatrix}
S - n^2 \cos^2\theta & -iD & n^2 \sin\theta \cos\theta \\
iD & S - n^2 & 0 \\
n^2 \sin\theta \cos\theta & 0 & P - n^2 \sin^2\theta
\end{bmatrix}, 
\end{equation}
where $n \equiv {c {k}}/{\omega}$, the wave vector is chosen as $\boldsymbol{k} = (k \sin\theta, 0, k \cos\theta)$.
In the cold plasma limit without the drift-kinetic approximation, we refer as ``full-CPDR", the dielectric tensor coefficients are given by \cite{stix1992waves}

\begin{equation} \label{eq:FK}
S = 1 - \sum_s \frac{\omega_{ps}^2}{\omega^2 - \Omega_{cs}^2}, ~~~
D = \sum_s \frac{\Omega_{cs} \omega_{ps}^2}{\omega (\omega^2 - \Omega_{cs}^2)},
~~~
P = 1 - \sum_s \frac{\omega_{ps}^2}{\omega^2}.
\end{equation}
The resulting dispersion relation usually has five sets of roots. For fixed parameters, each wavevector $k$ corresponds to five eigenmodes, corresponding to the fundamental cold two-fluid plasma wave branches. Since the characteristics and nomenclature of these ``full-CPDR" branches depend on the wavevector and propagation angle, detailed descriptions of these different modes can be found in \cite{stix1992waves, chen1987waves}.

For the hybrid model with drift-kinetic electrons and fully kinetic ions, retaining both the displacement current and the electron polarization current, we refer as DeFi, the coefficients become
\begin{equation}\label{eq:DeFi}
S = 1 + \frac{\omega_{pe}^2}{\Omega_{ce}^2} - \frac{\omega_{pi}^2}{\omega^2 - \Omega_{ci}^2}, 
~~~
D = \frac{\omega_{pe}^2}{\omega |\Omega_{ce}|}+\frac{\Omega_{ci} \omega_{pi}^2}{\omega (\omega^2 - \Omega_{ci}^2)},
~~~
P = 1 - \frac{\omega_{p}^2}{\omega^2}.
\end{equation}
The DeFi dispersion relation has four sets of roots.

When the displacement current is removed and electron polarization (EP) current is included explicitly, which we refer to as DeFi-QN-EP, the coefficients reduce to
\begin{equation}\label{eq:DeFi-Epolar}
S =  \frac{\omega_{pe}^2}{\Omega_{ce}^2}  - \frac{\omega_{pi}^2}{\omega^2 - \Omega_{ci}^2}, 
~~~
D = \frac{\omega_{pe}^2}{\omega |\Omega_{ce}|}+\frac{\Omega_{ci} \omega_{pi}^2}{\omega (\omega^2 - \Omega_{ci}^2)},
~~~
P =  - \frac{\omega_{p}^2}{\omega^2}.
\end{equation}
The dispersion relation of DeFi-QN-EP has three sets of roots.

Finally, in the DeFi-QN model, where both the displacement current and the electron polarization current are neglected, the coefficients take the form
\begin{equation}\label{eq:DeFi-QN}
S =    - \frac{\omega_{pi}^2}{\omega^2 - \Omega_{ci}^2}, 
~~~
D = \frac{\omega_{pe}^2}{\omega |\Omega_{ce}|}+\frac{\Omega_{ci} \omega_{pi}^2}{\omega (\omega^2 - \Omega_{ci}^2)},
~~~
P =  - \frac{\omega_{p}^2}{\omega^2},
\end{equation}
The DeFi-QN has 2 sets of roots.

To clearly illustrate the wave solutions, we adopt a reduced mass ratio $m_i/m_e=4$ and a moderate density parameter $\omega_{pe}/|\Omega_{ce}|=1$. The dispersion branches for parallel and perpendicular wave propagation are shown in Fig.~\ref{fig:compare_hybrid_CPDR}. Figure~\ref{fig:compare_hybrid_CPDR} compares the wave branches obtained from the DeFi model, the DeFi-QN-EP model, and the DeFi-QN model with the ``full-CPDR" solutions.
As shown in Figs.~\ref{fig:DeFi_kpar} and \ref{fig:DeFi_kperp}, when both the displacement current and the electron polarization current are retained, high frequency electromagnetic modes-including the L and X branches, plasma oscillation (Langmuir wave)-are recovered, although some with modified frequencies compared to the ``full-CPDR" branches \cite{Meng2025}.

As shown in Figs.~\ref{fig:DeFi-QN-EP_kpar} and \ref{fig:DeFi_QN_kperp}, in the DeFi-QN-EP model, high-frequency L and O branches are still present. These modes impose severe CFL constraints and stringent time step restrictions, which are unfavorable for efficient time integration. In particular, the L and O modes remain high frequency as $k \to 0$. Moreover, as shown in Fig.~\ref{fig:DeFi-QN-EP_kperp}, branch~2 CAW-LHW does not accurately reproduce the lower-hybrid  frequency at high $k$ due to the absence of the displacement current.  

In contrast, the quasi-neutral formulation without electron polarization current DeFi-QN model completely removes these high-frequency branches, retaining only the two low-frequency modes. The corresponding frequencies vanish as $k \to 0$, yielding a much more favorable spectral structure for explicit time stepping. 

However, the remaining right-handed polarized branch (labeled with 2: R CAW-WHW)  in the DeFi-QN model follows a whistler-like dispersion relation at high $k$, as demonstrated in Figs.~\ref{fig:DeFi_QN_kpara} and \ref{fig:DeFi-QN_theta60}, where the numerical results closely match the analytical scaling $\omega \propto k_\parallel^2$. This agreement indicates that the DeFi-QN model can captures the CAW to the WHW physical wave behavior in the small-$k$ regime.

At the same time, the high-frequency nature of this R-wave branch imposes stringent numerical stability constraints and significantly restricts the allowable time step. This observation further motivates the use of implicit \cite{YChen2009VlasovDrift} or implicit-explicit time-integration schemes for advancing the electromagnetic fields.
\subsection{Validity regime and high-$k$ behavior of the DeFi-QN model}
In the DeFi-QN model, the quasi-neutral approximation removes the Langmuir wave and the electron cyclotron resonance. As a result, the electromagnetic spectrum is reduced to low-frequency branches that are relevant to Alfv\'enic and Hall-driven dynamics.

\begin{figure}[htbp]
    \centering
    \includegraphics[width=0.5\linewidth]{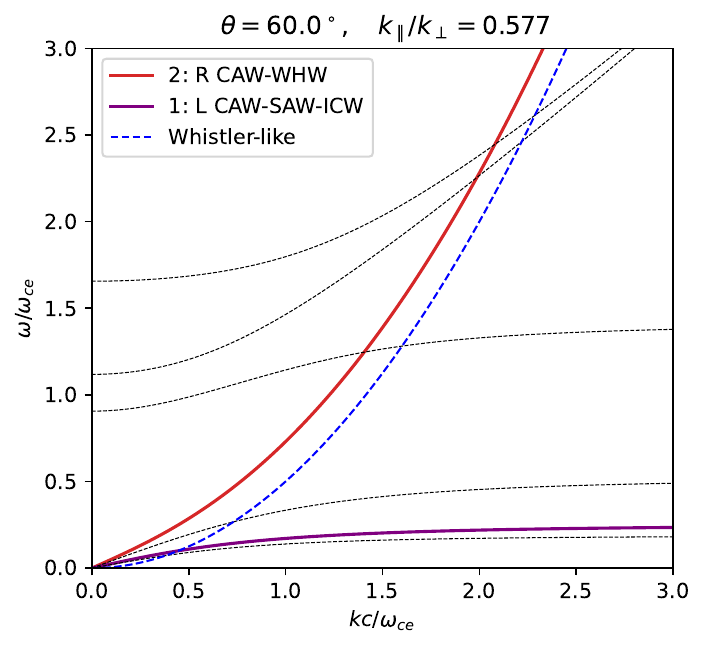}
    \caption{DeFi-QN model for oblique propagation with $\theta=60^\circ$.
The R-wave branch is compared with the whistler-like dispersion relation
$\omega = |\Omega_{ce}| (c^2/\omega_{pe}^2)\, k k_\parallel$. }
    \label{fig:DeFi-QN_theta60}
\end{figure}

This dispersion relation admits two electromagnetic branches, as shown in Figs.~\ref{fig:DeFi_QN_kpara}, \ref{fig:DeFi_QN_kperp} and \ref{fig:DeFi-QN_theta60}. The upper branch is right-handed polarization and the lower branch is left-handed polarization.
The upper branch scales quadratically with $k$, leading to severe timestep restrictions in numerical simulations. This high-frequency branch can be entirely removed by using an electrostatic model. For parallel propagation ($k = k_\parallel$), the right-hand branch satisfies
\begin{equation}
    \frac{c^2k^2}{\omega^2}= R = S+D=\frac{\omega_{pe}^2}{\omega |\Omega_{ce}|} - \frac{\omega_{pi}^2}{\omega (\omega+ \Omega_{ci})}. \label{eq:R_Whistler}
\end{equation}
Retaining the positive eigenfrequency, we obtain the upper (right-hand) branch
\begin{equation}
    \omega_R = \frac{|\Omega_{ce}|ck}{2\omega_{pe}^2} \left(c k + \sqrt{c^2 k^2 + 4 \omega_{pi}^2}\right).\label{eq:wR_upperbranch_kpar}
\end{equation}
By following the same procedure, the lower (left-hand) branch is
\begin{equation}
\frac{c^2k^2}{\omega^2}= L=S-D,  \qquad \Rightarrow
\omega_L =  \frac{|\Omega_{ce}| c k}{2\omega_{pe}^2}  \left( -c k + \sqrt{c^2 k^2 + 4 \omega_{pi}^2} \right).\label{eq:wL_lowerbranch_kpar}
\end{equation}
In the long-wavelength limit $k \to 0$, both branches degenerate to,
\begin{equation}
    \omega_R = \omega_L
    \approx
    V_{A,i} k \; \;,
\end{equation}
recovering the Alfv\'en wave dispersion relation.

If we neglect the ion term ($\omega_{pi}^2=0$) in Eq. \eqref{eq:wR_upperbranch_kpar}, the $\omega_R$ reduces to the classical whistler dispersion relation 
\begin{equation}
    \omega_{WH}=c^2k^2|\Omega_{ce}|/\omega_{pe}^2.
\end{equation}
In the low-$k$ regime, where $\omega \ll |\Omega_{ce}|$ and $\omega \ll \omega_{LH}$ (low hybrid wave frequency $\omega_{LH}\simeq \sqrt{\Omega_{ci}|\Omega_{ce}|}$),
the upper branch continuously connects the compressional Alfv\'en wave (CAW) to the whistler wave (WHW). In this regime, the mode is physical and correctly captured by the DeFi-QN model.
Figure~\ref{fig:DeFi-QN_theta60} confirms that the numerical solution follows the expected quadratic scaling
$\omega \propto k_\parallel^2$.
At larger $k$, however, the absence of electron cyclotron and lower-hybrid resonances causes the R-wave branch to become unbounded.
This divergence does not signify a physical instability, but reflects the breakdown of the drift-kinetic and quasi-neutral orderings. For perpendicular propagation ($k=k_\perp$), the eigenfrequency is $\omega_R = V_{A,i} k$, which corresponds to the Alfv\'en wave dispersion relation.

In summary, the valid regime of the DeFi-QN model requires $\omega \ll |\Omega_{ce}|$, and $\omega \ll \omega_{LH}\simeq \sqrt{\Omega_{ci}|\Omega_{ce}|}\sim \sqrt{m_i/m_e}\Omega_{ci} $. The drift-kinetic model for electrons requires $\omega \ll |\Omega_{ce}|$,
so electron gyro-motion can be averaged and electron polarization effects remain negligible.
The electron cyclotron motion removed so R-mode does not encounter $|\Omega_{ce}|$ resonance and becomes unbounded for increasing $\kpar$ as shown in Fig. \ref{fig:DeFi_QN_kpara}. The drift-kinetic ordering further requires
$\kperp r_{Le}\ll 1$, where $r_{Le}$ is the electron Larmor radius. If $\kperp r_{Le} \gtrsim 1$, electron Finite Larmor Radius (FLR) effects cannot be neglected. 
And quasi-neutrality imposes $k\lambda_{De} \ll 1$, where $\lambda_{De}=\sqrt{\frac{\epsilon_0 T_e}{n_e e^2}}=v_{te}/\omega_{pe}$, which eliminates Langmuir and electron plasma oscillations.
In particular the lower-hybrid (LH) resonance requires electron polarization to couple ion cyclotron motion to the electron response; removing electron polarization prevents this coupling, so the DeFi-QN R-mode does not encounter the LH resonance and becomes unbounded for increasing $\kperp$ as shown in Fig. \ref{fig:DeFi_QN_kperp}.
Figure~\ref{fig:DeFi-QN_theta60} shows that when $\kperp$ is comparable to $\kpar$,
the L-wave branch in the DeFi-QN model exhibits a small frequency deviation from the  branch of ``full-CPDR" as
$k$ increases. This suggests that the DeFi-QN model provides a reliable description in the low-$k$ regime and for small to moderate $\kperp/\kpar$.

\section{Numerical results}\label{sec:benchmark}
The DeFi-QN model has been implemented in the \texttt{GEMPICX} code, which is used for all simulations reported in this work.
We simplify the model for the case of a slab with a constant and uniform magnetic field $\Bex$ and consider only $\mu=0$ for drift-kinetic electrons, i.e., $T_{e\perp}=0$. Under these conditions, the electron magnetization $\mathbf{M}_e$ vanishes.

For clarity, we summarize some commonly used plasma parameters in this section and their relationships:  the thermal velocity is defined as $v_{ts}\equiv \sqrt{T_s/m_s}$, the sound speed $c_{\rm s}=\sqrt{T_e/m_i}$, and $\rho_{\rm sound}=c_{\rm s}/\omega_{ci}$, the electron beta $\beta_e=2\mu_0 n_0 T_e/\Bzex^2$, and their relation to the ion Alfv\'en speed  $ {c_{\rm s}}/{V_{A,i}}=\sqrt{\beta_e/2}$. Unless otherwise specified, a real mass ratio $m_i/m_e = 1836$ and density parameter $\omega_{pe}/|\Omega_{ce}| = 1$ are used.

The simulations employ a two-dimensional configuration space ($N_x \times N_y$ grid), while particle velocities and electromagnetic fields retain all three Cartesian components. The simulation domain length is set according to the prescribed wavevector ${k}_{\text{input}}$ as $L = 2\pi/k_{\text{input}}$. Third-order B-spline particle shapes are used in all spatial directions. Particle positions and velocities are initialized using a low-rank Sobol sampler to reduce initial loading noise. Periodic boundary conditions are applied in all directions.

Time integration is performed using either a low-storage Runge-Kutta (LSRK) scheme or an IMEX splitting scheme (referred as ``RK-CN-RK"): the explicit terms are advanced with a half-step LSRK scheme, the stiff Hall term in the magnetic field equation Eq.~\eqref{eq:B_implicit} is integrated using a full-step Crank-Nicolson scheme, and the remaining explicit terms are completed with a final half-step LSRK. Depending on the desired accuracy, third-order (3-stage) or fourth-order (5-stage) LSRK can be employed \cite{Meng2025}.
 
Note that we did not apply any Fourier filtering or smoothing in the simulations presented later in this paper, so all wavenumbers $\bk$ are present in the results.

\subsection{\texorpdfstring{$k_\perp$}{k_perp} waves}

For perpendicular propagation, the background magnetic field is set to $\Bex = (0, 0, 1)$ along the $z$-direction. The $k_{\rm input} = [0.005,\, 0.01]$, giving a domain
\[
[0,\, 2\pi/0.005]\, d_e \;\times\; [0,\, 2\pi/0.01]\, d_e,
\]
using a grid of $256 \times 8$ cells. Here $d_e = c/\omega_{pe}$ denotes the electron inertial length.

\begin{figure}[htbp]
    \centering
    \includegraphics[width=0.5\linewidth]{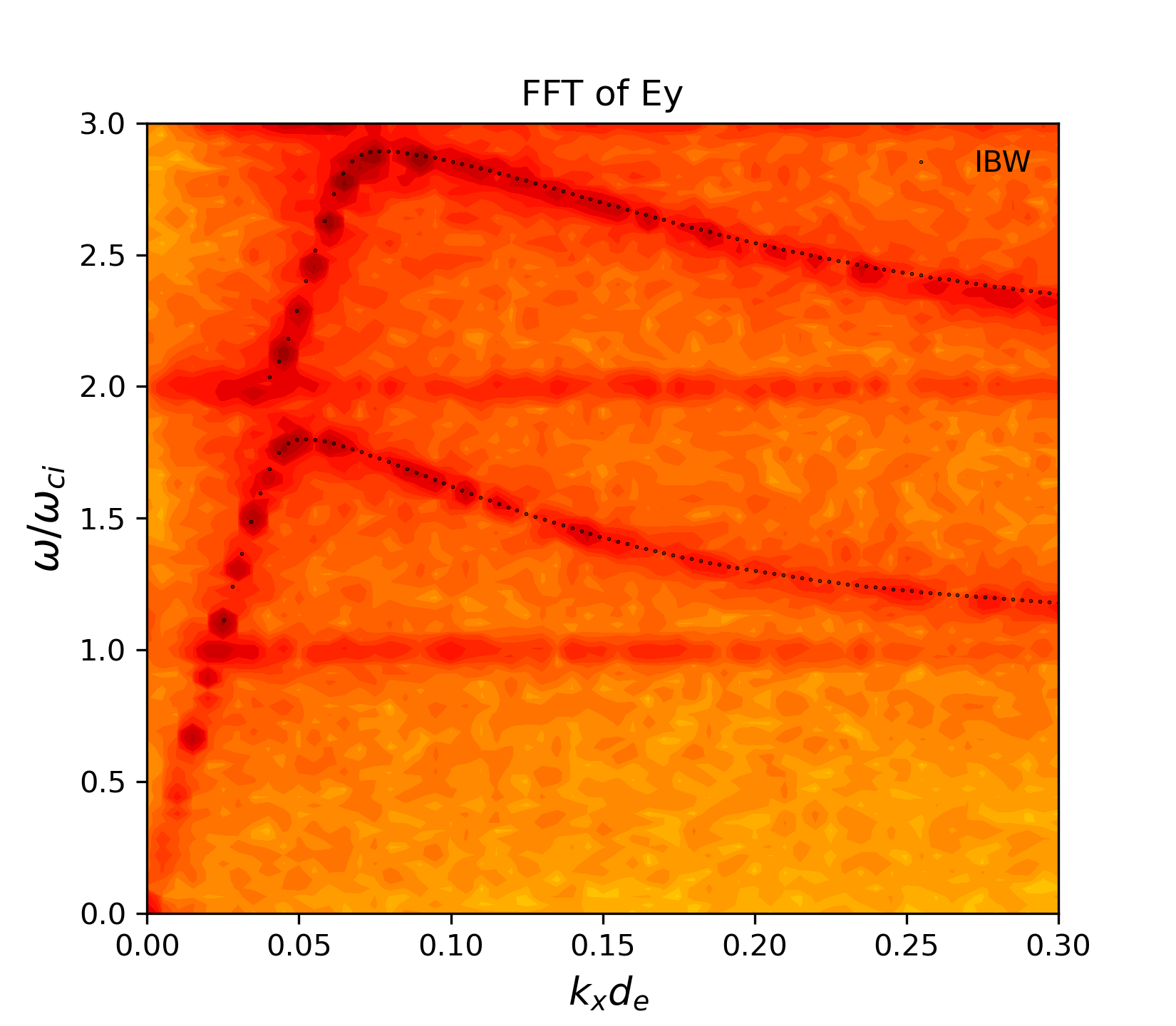}
    \caption{Perpendicular wave spectrum from the DeFi-QN simulation. $E_y$ is averaged over $x$ and $z$ and plotted versus $k_x$ ($\kperp$). Black dots indicate the theoretical ion Bernstein wave frequencies. The simulation captures the ion cyclotron frequency and higher-order Bernstein modes. $m_i=1836$,\; $v_{te}=0.2, \; T_i/T_e=1$}
    \label{fig:kperp_mi1836}
\end{figure}

Figure~\ref{fig:kperp_mi1836} presents the spectrum of $E_y$, averaged over $x$ and $z$, plotted versus $k_x$ ($k_\perp$).  The perturbation $E_y$ is perpendicular to both $\Bex$ and the wave vector.
Simulations use $200$ particles per cell per species. A time step  \( \Delta t = 10 \, \omega_{ce}^{-1} \), and the total simulation time  \( T = 200 \, \omega_{ci}^{-1} \) are used. The simulation parameters are $m_i = 1836$, $v_{te} = 0.2$, and $T_i/T_e = 1$.

The black dots indicate the ion Bernstein wave (IBW) frequencies predicted by kinetic theory. The simulation results show excellent agreement with the theoretical predictions, including a clear identification of the ion cyclotron frequency. 
Small deviations at higher $k$ can be attributed to finite grid resolution and particle noise. Overall, these results confirm that the DeFi-QN model accurately captures perpendicular wave physics in magnetized plasmas under the considered parameter regime.

\subsection{\texorpdfstring{$k_\parallel$}{k_parallel} transverse waves}

To investigate parallel-propagating transverse waves, the background magnetic field is set to $\Bex = (1, 0, 0)$ along the $x$-direction.  We use $k_{\text{input}} = [0.002,\, 0.01]$, corresponding to a simulation domain
\[
[0,\, 2\pi/0.002]\, d_e \;\times\; [0,\, 2\pi/0.01]\, d_e,
\]
again discretized with a grid of $256 \times 8$ cells.  The time step  is \( \Delta t = 10 \, \omega_{ce}^{-1} \) and the total simulation time  \( T = 100 \, \omega_{ci}^{-1} \). Each species is represented by $200$ particles per cell. Additional parameters are $v_{te} = 0.1$, and $T_i/T_e = 1$. The simulation is initialized with discrete particle noise, from which the plasma eigenmodes naturally emerge as the system evolves.

\begin{figure}
     \centering
     \includegraphics[width=0.5\linewidth]{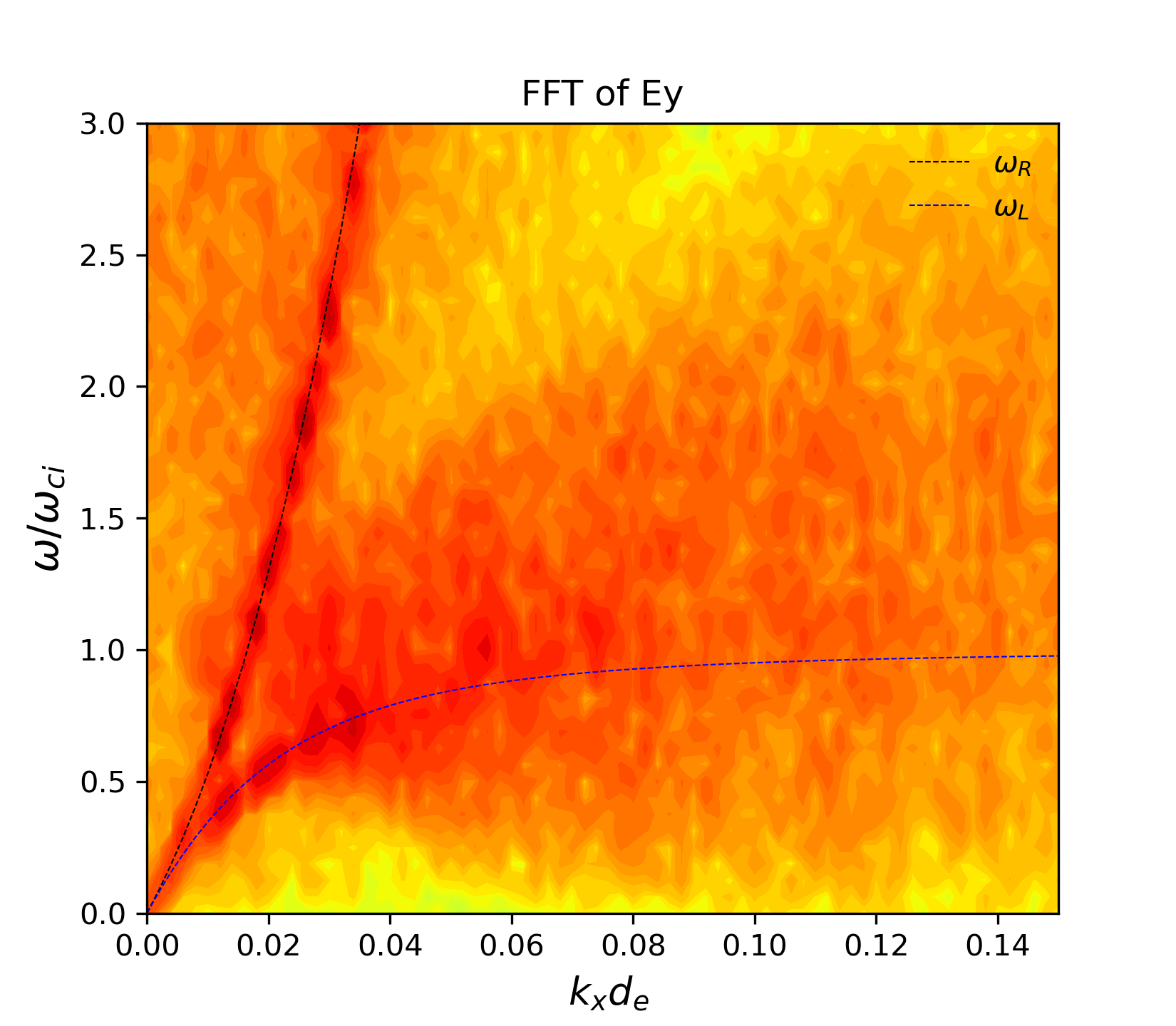}
     \caption{Parallel wave spectrum from the DeFi-QN simulation with $m_i=1836$,\; $v_{te}=0.1,\; T_i/T_e=1$. The power spectrum of $E_y$ is shown versus $k_x$ ($k_\parallel$). Dashed lines indicate theoretical right- and left-hand polarized branches. The cone-like feature near $\omega_{ci}$ corresponds to damped higher-order Alfv\'en-cyclotron modes. }
     \label{fig:kpar_mi1836}
 \end{figure}

Figure \ref{fig:kpar_mi1836} shows the simulation results for the perpendicular waves. The $E_y$ is averaged over $x$ and $z$. The spectrum of $E_y$ is plotted versus $k_x$. Here $k_x$ corresponds to the parallel wave number $\kpar$, and the perturbation $E_y$ is perpendicular to the background magnetic field $\Bex$.

The measured spectrum agrees well with the theoretical right- and left-hand branches given by Eqs.~\eqref{eq:wR_upperbranch_kpar} and \eqref{eq:wL_lowerbranch_kpar}. A cone-shaped structure emanating from ($k$, $\omega$) = (0, $\omega_{ci}$) on the spectrum is also visible and corresponds to  the heavily damped higher-order Alfv\'en-cyclotron modes \cite{araneda2012interactions}. 

\subsection{Shear and compressional Alfv\'en waves}
The ``cancellation problem'' is a well-known numerical issue in traditional electromagnetic gyrokinetic models formulated in terms of the scalar potential $\phi$ and vector potential $\mathbf{A}$ \cite{cummings1994gyrokinetic}. A critical parameter is the product of the electron skin depth and the perpendicular wave vector $\kperp d_e$. When  $1/(\kperp d_e)>1$, the ``cancellation problem'' becomes severe, particularly when using the pure Hamiltonian ($p_\parallel$) form. The mixed-variable pullback scheme can resolve the ``cancellation problem'' \cite{mishchenko2014pullback}. It has been shown that the mixed-variable pullback scheme is efficient in mitigating the ``cancellation problem'' and achieving high accuracy, and mitigating the noise level \cite{lu2025generalized}. In addition to the $\mathbf{A}-\phi$  model, it has been previously reported that the GK-E\&B model can be applied to simulation of shear Alfv\'en wave in the small electron skin depth limit but the ``cancellation problem'' also exists in the $E_\|$ equation, although it is not exactly the same equation as that in the $\mathbf{A}$ and $\phi$ models \cite{rosen2022EB}. The implicit scheme has been developed to simulate the shear Alfv\'en waves without the same ``cancellation problem'' as that in the Hamiltonian form \cite{lu2021implicit}. 
Furthermore, a mixed full-$f$-$\delta f$ method using the mixed variable-pullback scheme has been developed, and shown to perform well in the small electron skin depth regime \cite{lu2023full}.  

\begin{figure}
    \centering
    \includegraphics[width=0.5\linewidth]{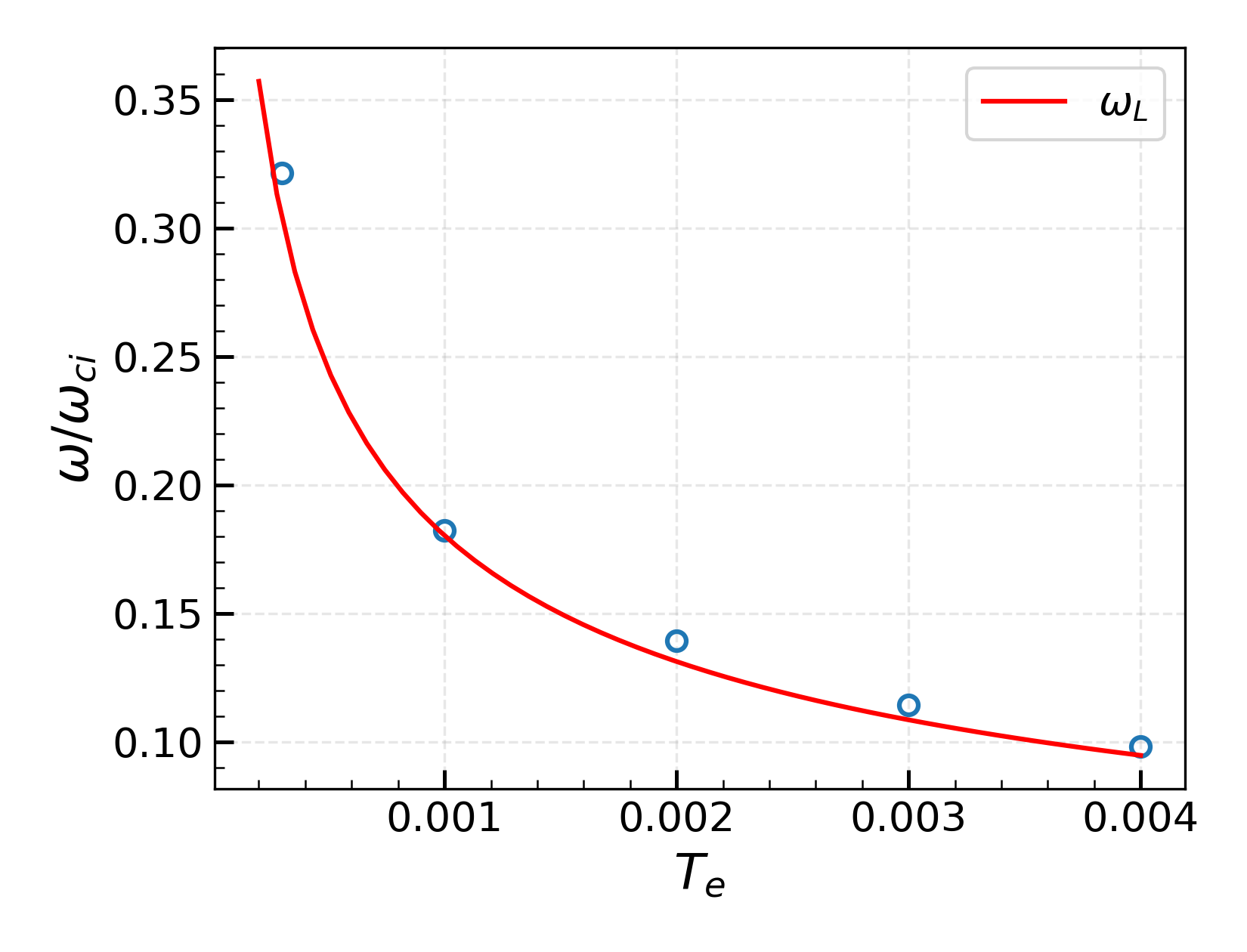}
    \caption{Simulated shear Alfv\'en wave frequencies vs  $T_e$ (dots)  compared with the dispersion relation (solid line) for $\kperp=0$. The $\omega_L$ is calculated from \eqref{eq:wL_lowerbranch_kpar}.}
    \label{fig:SAE}
\end{figure}

To benchmark shear Alfv\'en eigenmodes (SAEs), we adopt parameters similar to those in \cite{YChen2009VlasovDrift}, with $\kpar \rho_{\rm sound} = 0.0063$ and $\kperp = 0$. Here we use $T_e = T_i$, where the electron temperature $T_e$ is varied from $0.0003$ to $0.004$. For finite $\kpar$, two wave branches exist, corresponding to the right- and left-handed polarized modes. The SAE corresponds to the left-handed branch. In Fig.~\ref{fig:SAE}, we compare the simulated results with the analytical dispersion relation. Notably, this case implies, $1/(k_\perp d_e) \to \infty$, which is extremely challenging for the traditional $p_\|$ formula using the $\mathbf{A}$ and $\phi$ model. However, the $\mathbf{E}$-$\mathbf{B}$ model can handle this parameter regime, demonstrating a different and good feature compared with traditional $\mathbf{A}$-$\phi$ schemes.

\begin{figure}[htbp]
    \centering
    \includegraphics[width=0.5\linewidth]{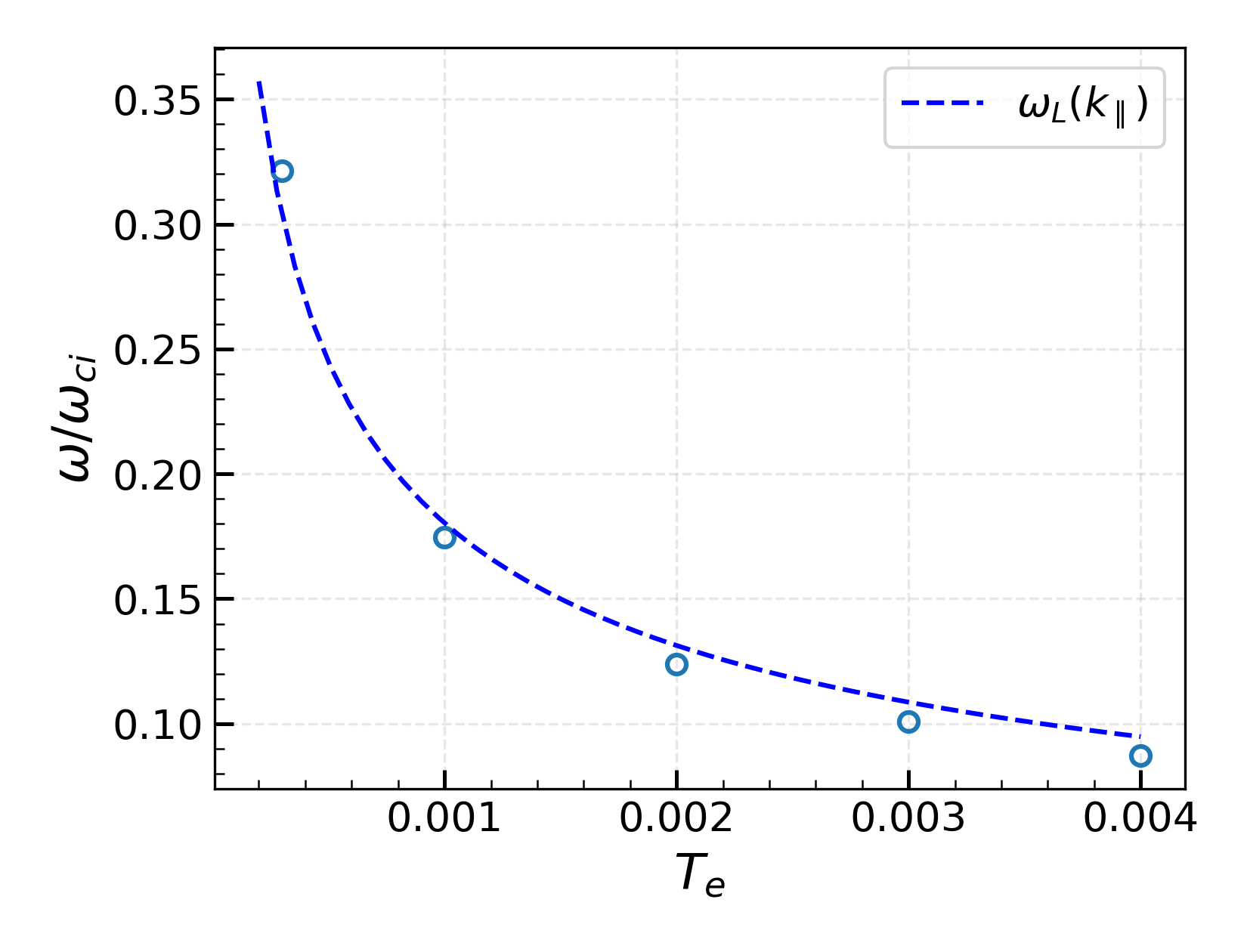}
    \caption{Simulated shear Alfv\'en wave frequencies vs  $T_e$ (dots) compared with the dispersion relation (solid line) for  $1/(k_\perp d_e)=100$. The $\omega_L$ is calculated from \eqref{eq:wL_lowerbranch_kpar}. The energy evolution of $T_e=0.003$ case is shown in Fig.~\ref{fig:SAE_energy}.}
    \label{fig:SAE_kperp0.01}
\end{figure}

Another simulation of the SAE was performed with same $\kpar \rho_{\rm sound} = 0.0063$ and a large value of $1/(k_\perp d_e)=100$ ($k_y=0.01$). In the $\mathbf{A}$-$\phi$ model, the corresponding ratio between the physical part and the adiabatic part in the parallel Amp\`ere's law is $|\delta A_\|/d_e|^2/|k_\perp\delta A_\||^2=10^4$, which can lead to a severe ``cancellation problem''. The frequencies of the SAE for different values of $T_e$ are shown in Fig.~\ref{fig:SAE_kperp0.01}. 

\begin{figure}[htbp]
    \centering
    \includegraphics[width=0.95\linewidth]{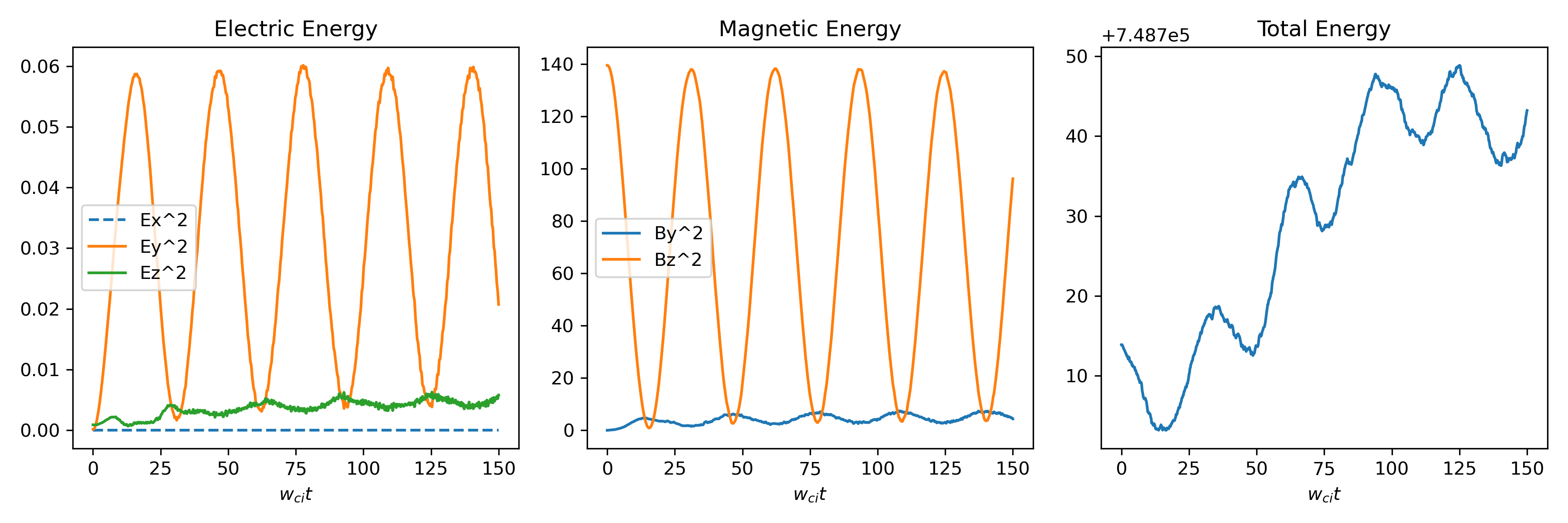}
    \caption{Energy evolution of the $T_e=0.003$, $k_x \rho_{\rm sound} = 0.0063$, and $k_y=0.01$ case. With $\Bex \parallel \hat{\mathbf{x}}$, the parallel electric field energy $E_x^2$ is visibly zero.  Fluctuations are dominated by  $E_y^2$ and $B_z^2$, exhibiting typical SAE polarization.}
    \label{fig:SAE_energy}
\end{figure}

As shown in Fig.~\ref{fig:SAE_energy}, energy evolution for $T_e=0.003$ and $k_y=0.01$ shows typical shear Alfv\'enic polarization. With $\Bex \parallel \hat{\mathbf{x}}$ and $\mathbf{k}_{\perp} \parallel \hat{\mathbf{y}}$, the fluctuations are dominated by fields energy $E_y^2$ and $B_z^2$, the latter representing transverse magnetic field line bending. Parallel components $E_x$ and $B_x$ remain negligible, consistent with the incompressible nature of the mode. The SAW displays stable oscillations with negligible growth or damping over the duration of $150 \Omega_{ci}^{-1}$. Minor fluctuations observed in $E_z^2$ and $B_y^2$ are attributed to residual PIC noise and the absence of spectral filtering, which precludes a perfectly pure left-hand polarized state. Furthermore, total energy is well-conserved; the ratio of total energy fluctuations is maintained at a minimal level ($\sim 6.6 \times 10^{-5}$), demonstrating the numerical stability of the simulation.  These results confirm the capability of the present work in simulating SAEs in the limit of small electron skin depth.


\begin{figure}
    \centering
    \includegraphics[width=0.5\linewidth]{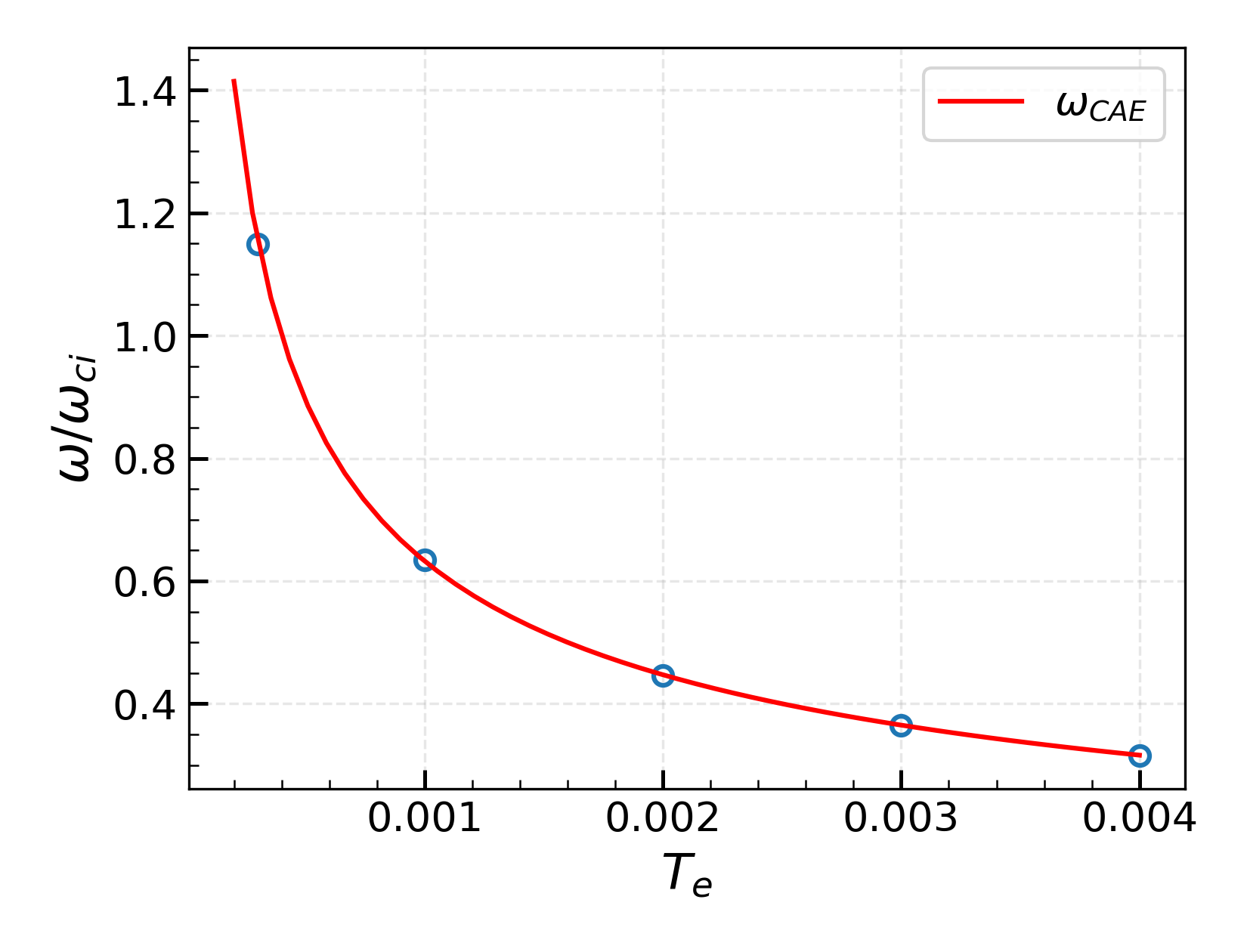}
    \caption{Simulated compressional Alfv\'en wave frequencies vs  $T_e$ (dots)  compared with the dispersion relation (solid line).}
    \label{fig:CAE}
\end{figure}

For compressional Alfv\'en eigenmodes (CAEs), we again follow the setup of \cite{YChen2009VlasovDrift}, using $\kperp\rho_{\rm sound}=0.02$ and $\kpar=0$. In the long wavelength limit, the CAE frequency is given by $\omega_{\rm CAE}=kV_{A,i}$. We use $\Bex=[0,0,1]$, $64\times8$ cells, domain with $k_{\rm input}=[0.02/\rho_{\rm sound},0.01]$, $\Delta t = 10$. The simulations are initialized with $B_z=0.05\cos(k_xx)$. 
Figure~\ref{fig:CAE} shows the simulated frequencies as a function of $T_e$, compared with the analytical prediction.

Overall, the model demonstrates good agreement with theoretical predictions for both shear and compressional Alfv\'en waves, indicating its robustness and accuracy in electromagnetic wave simulations. 
These AE results($k_\parallel/\kperp=\infty$), spectrum for $k_\parallel$, $\kperp$
show our model works well in the ``MHD limit".

\subsection{Longitudinal waves in the quasi-neutral model}
In the quasi-neutral model, high-frequency Langmuir modes are eliminated, and longitudinal dynamics are governed by low-frequency kinetic effects. For a Maxwellian distribution, we derive the susceptibility tensor for drift-kinetic electrons. The $zz$ component of $\boldsymbol{\chi}_e$ 
$$\boldsymbol{\chi}_{e,zz}= \frac{I_1}{\epsilon_0 \omega}$$
where \begin{align}
	{I_1(\omega, \boldsymbol{k})} &=  \frac{q_e^2 n_M \Bzex }{m_e} \int  \frac{\vpar \fracp{f_{eM}}{\vpar}}{\omega - \kpar \vpar} \dd \vpar \dd \mu
\end{align}
The $zz$ component depends only on $\partial f_{e0}/\partial \vpar$, which implies that temperature anisotropy ($T_{e,\perp} \neq T_{e,\parallel}$)  does not affect $\boldsymbol{\chi}_{e,zz}$ and therefore does not modify the parallel dispersion relation. Omitting the subscript $zz$, the parallel modes are governed by the kinetic dispersion relation without the quasi-neutral assumption \cite{Meng2025}
\begin{equation}\label{eq:orig_Dzz}
 \boldsymbol{\Lambda}_{zz}(\omega,k)=\epsilon_{zz}=1+\chi_e+\chi_i=0
\end{equation}
where the species susceptibilities are given by 
$$\chi_s = (\kpar^2 \lambda_{Ds}^2)^{-1}\!\left[1+\xi_s Z(\xi_s)\right]$$ with $\xi_s=\omega/(\kpar v_{ts}\sqrt{2})$, $v_{ts}=\sqrt{T_s/m_s}$, and $\lambda_{Ds}=v_{ts}/\omega_{ps}$.
In the quasi-neutral limit, this reduces to
\begin{equation}\label{eq:QN_Dzz}
    \boldsymbol{\Lambda}_{zz}(\omega,k)=\epsilon_{zz}=\chi_e+\chi_i=0 \;,
\end{equation}
 Using $\lambda_{Di}^2/\lambda_{De}^2=T_i/T_e$, the quasi-neutral dispersion relation can be renormalized as
\begin{equation}\label{eq:QN_Dzz_renorm}
    1+\xi_e Z(\xi_e)+\frac{T_e}{T_i}\left[1+\xi_i Z(\xi_i)\right]=0.
\end{equation}
In the quasi-neutral limit, all solutions are damped. However, for $T_i/T_e<<1$ and $m_e/m_i<<1$, a weakly damped root exists, corresponding to the ion acoustic wave (IAW). The IAW is a low-frequency longitudinal electrostatic mode with dispersion relation $\omega \approx c_{\rm s} k$, where $c_{\rm s} = \sqrt{T_e/m_i}$ is the ion sound speed. It arises from the balance between electron pressure (restoring force) and ion inertia, and is weakly damped in the regime $v_{ti} \ll \omega/k \ll v_{te}$. In a magnetized plasma, the IAW propagates primarily along the background magnetic field $\Bex$, since electrons must move freely along field lines to maintain pressure balance. The wave is dominated by quasi-neutral density perturbations ($\delta n_e \approx \delta n_i$) and is longitudinal, meaning that the perturbations, velocity response, and the electric field, are mainly parallel to $\Bex$. Magnetic fluctuations are negligible, so the IAW is essentially a compressive electrostatic sound wave of the plasma.

The IAW corresponds to  the least damped solution of Eq. \eqref{eq:QN_Dzz_renorm}. 
We adopt the parameter set $m_i/m_e = 200$, $T_i/T_e = 10^{-4}$, $v_{te} = 1$, and $k=2\pi/10$ which has been widely used in studies of energy-conserving and structure-preserving PIC methods \cite{chen2011energy,kormann2021energy}. 
A quick estimate of the ion acoustic wave (IAW) frequency and damping rate can be obtained in the cold-ion limit. In this approximation, the IAW frequency is given by $\omega = c_{\rm s} k$. In the long-wavelength regime ($k \ll 1/\lambda_{De}$) and for cold ions ($T_i \ll T_e$), the damping rate is small and can be approximated by
\[
\gamma_{\mathrm{IAW}} \approx \omega_{\mathrm{IAW}} \sqrt{\frac{\pi}{8}} \sqrt{\frac{m_e}{m_i}}.
\]

For the present parameter set, this approximation yields
\[
\omega/\omega_{ci} = 8.8858 - 0.39374\, i.
\]

\begin{figure}[htbp]
    \centering
    \includegraphics[width=0.5\linewidth]{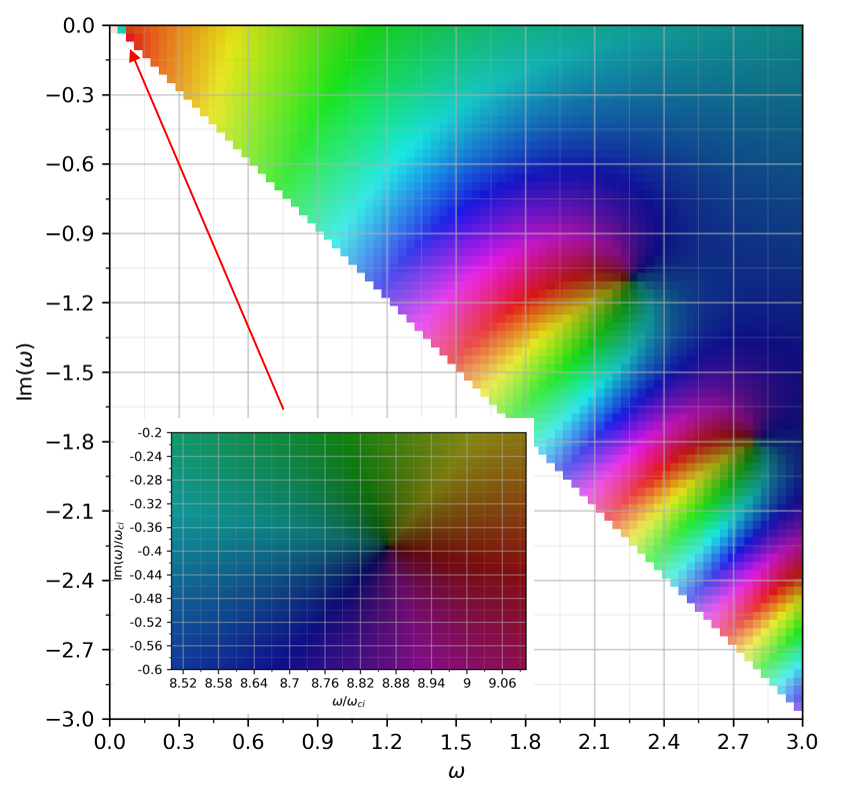}
    \caption{Dispersion relation given by Eq.~\eqref{eq:QN_Dzz_renorm} for $m_i/m_e = 200$, $T_i/T_e = 10^{-4}$, $v_{te} = 1$, and $k = 2\pi/10$. The magnitude $|\Lambda_{zz}(\omega,k)|$ and its phase are shown in the complex $\omega$ plane, where the hue represents $|\Lambda_{zz}|$ and the color encodes its phase. The eigenfrequency (root of $\Lambda_{zz}(\omega,k)=0$) corresponds to the location where is dark and the phase exhibits a sharp concentration (phase singularity). The IAW is the least damped mode, visible in the inset. The phase singularity near $\omega \approx 2.25$ corresponds to the second least damped root.}
    \label{fig:D_zz_k1}
\end{figure}
A more accurate determination of the IAW frequency and damping rate requires solving the full kinetic dispersion relation. The roots of Eq.~\eqref{eq:QN_Dzz_renorm} are shown in Fig.~\ref{fig:D_zz_k1}, where the IAW appears as the least-damped mode (see inset). While numerous damped roots exist, the higher-order modes exhibit significantly larger frequencies and damping rates, so the IAW dominates the plasma response. 

In quasi-neutral limit, higher-order modes $\omega_n$ scale exactly linearly with $k_\parallel$ just like the IAW, since  $\omega$ and $k_\parallel$ only appear together inside the argument $\xi_s$. Consequently, each mode satisfies $\omega_n=v_{ph,n} \cdot k$, where $v_{ph,n}$ is a complex number. For the higher-order modes, the phase velocity exceeds the electron thermal speed, i.e. $\mathrm{Re}(v_{ph,n})>v_{te}$.

Since IAWs are Landau-damped modes, the dispersion function involves large numbers that nearly cancel, e.g., in terms of the form $1+\xi Z(\xi)$, leading to round off errors and instability in contour-integration-based evaluations. As also shown in the inset of Fig.~\ref{fig:D_zz_k1}, the magnitude $|\boldsymbol{\Lambda}_{zz}(\omega,k)|$ (the hue) varies only slowly near the IAW root.
In addition, in the quasi-neutral limit $k\lambda_D \ll 1$, the susceptibilities contain large prefactors $(k^2 \lambda_D^2)^{-1}$, further amplifying numerical errors. To improve numerical stability, we solve Eq.~\eqref{eq:QN_Dzz_renorm}, obtained by multiplying Eq.~\eqref{eq:QN_Dzz} by $k^2\lambda_{De}^2$, which removes the large factor and reduces cancellation errors. The equation is then solved using a Newton method, with the above analytical estimate as the initial guess. 

The resulting IAW frequency for the given parameter set ($T_e/T_i = 10^4$, $m_i/m_e = 200$), the fundamental IAW frequency calculated without the quasi-neutral condition is
\[
\omega_0/\omega_{ci}= 7.5215 - 0.20242 i.
\]
Under the quasi-neutral assumption, the frequency is 
\[
\omega_0/\omega_{ci}= 8.8658 - 0.39248 i.
\]
The second least damped root (the first higher-order mode) is found at  $\omega_1/\omega_{ci} = 452.52 - 217.8 i$ as shown in Fig. \ref{fig:D_zz_k1}. This mode is characterized by heavy damping, with a decay-to-frequency ratio of $\gamma/\omega_r=-0.4813$.

\subsection{Spectrum of longitudinal Spontaneous Fluctuations }
The spectral distribution of the electric field fluctuations in a magnetized, isothermal ($T_e=T_i=T$), and homogeneous plasma can be expressed via the standard fluctuation-dissipation theorem as \cite{sitenko1967, araneda2012interactions}
\begin{equation}
\langle E_i E_j \rangle_{\omega \bk}
=
4\pi i \frac{k_B T}{\omega}
\left[
(\boldsymbol{\Lambda}^{-1})_{ji}
-
(\boldsymbol{\Lambda}^{-1})_{ij}^*
\right],
\end{equation}
where $T$ is the plasma temperature and $\boldsymbol{\Lambda} ^{-1}$ is the inverse of the dispersion tensor.
Consider the dispersion tensor in a magnetized plasma for parallel propagation, $k=k_z$, setting $i=j=z$ and it yields
\begin{equation}
\langle E_z^2 \rangle_{\omega }
=
-8\pi
\frac{k_B T}{\omega}
\mathrm{Im}
\left(
\frac{1}{\epsilon_{zz}}
\right)
=
8\pi
\frac{k_B T}{\omega}
\frac{\mathrm{Im}(\epsilon_{zz})}
{|\epsilon_{zz}|^2}.
\end{equation}

This result shows that the longitudinal electric-field fluctuations are entirely determined by the longitudinal dielectric function $\epsilon_{zz}$.
For a non-isothermal plasma ($T_e\neq T_i$), which may be treated as a quasi-equilibrium system, the spectral density of the longitudinal electric field fluctuations is given by\cite{sitenko1967}
\begin{equation} \label{eq:multiTemp_fluc}
\langle E_z^2 \rangle_{\omega }
=
8\pi
\frac{k_B}{\omega}\sum_s
\frac{T_s \mathrm{Im}(\chi_{s})}
{|\epsilon_{zz}|^2}.
\end{equation}
the total dielectric function satisfies $\mathrm{Im}(\epsilon_{zz})=\sum_s
\mathrm{Im}(\chi_{s})$.

\begin{figure}[htbp]
    \centering
    \includegraphics[width=0.5\linewidth]{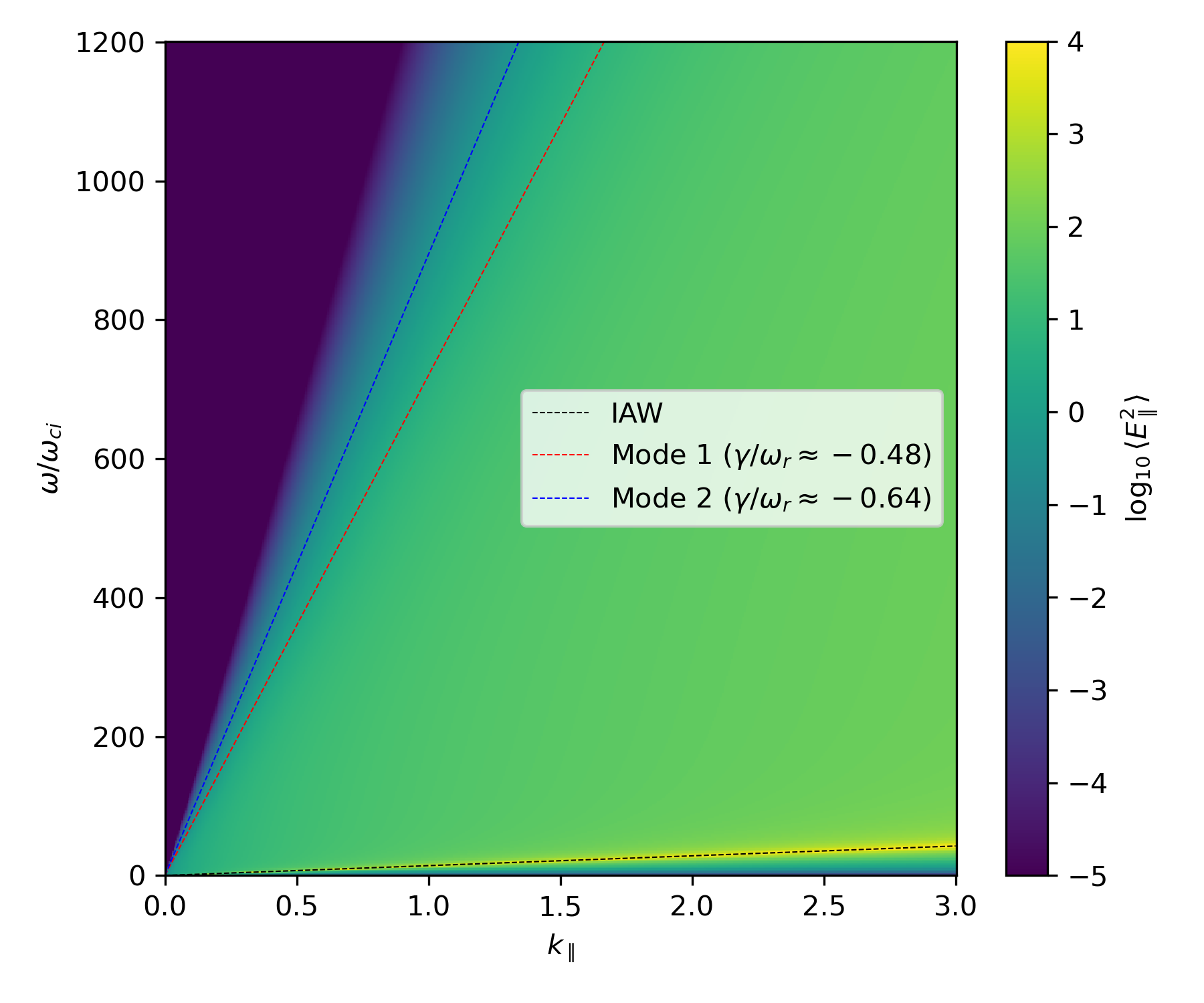}
    \caption{Analytical spectral density of the longitudinal electric field fluctuations $\langle E_\parallel^2 \rangle$ derived from Eq. \eqref{eq:multiTemp_fluc} for the quasi-neutral model. Parameters are set to $m_i/m_e = 200$, $T_i/T_e = 10^{-4}$, $v_{te} = 1, \omega_{pe}=1$. The theoretical dispersion branches for the ion-acoustic wave (IAW) and the higher-order damped modes are overlaid for comparison.  }
    \label{fig:ana_Ex_kx}
\end{figure}

By applying the quasi-neutral limit (substituting Eq. \eqref{eq:QN_Dzz} into Eq. \eqref{eq:multiTemp_fluc}), we obtain the analytical fluctuation spectrum for parallel propagation $(E_\parallel, k_\parallel)$,  as shown in Fig.~\ref{fig:ana_Ex_kx}. 
The fluctuation spectrum exhibits an intense, narrow ridge along the IAW frequency. Since the spectral power is proportional to $1/|\epsilon_{zz}|^2$, the  brightness of the spectrum traces the roots of the dispersion relation, where the plasma's collective response is most easily excited by thermal noise. The sharpness of the IAW peak is proportional to the ratio $T_e/T_i$. The IAW peak is particularly sharp at lower $k$, where Landau damping is minimal.
The observed triangular region of enhanced fluctuations at low frequencies corresponds to the transition between individual particle noise and collective behavior; in this regime, the spectral intensity is effectively bounded by the heavily damped high-order branches of the dispersion function, which constrain the emergence of fluctuations to the relatively ``free" zones of the $(\omega, k)$ plane \cite{araneda2012interactions}. 

\begin{figure}[htbp]
    \centering
    \includegraphics[width=0.45\linewidth]{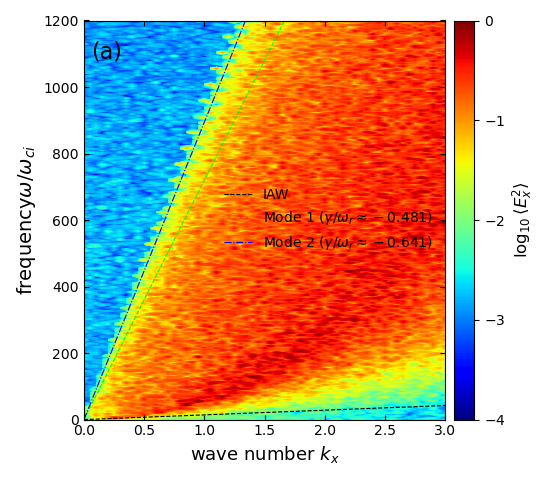}
    \includegraphics[width=0.45\linewidth]{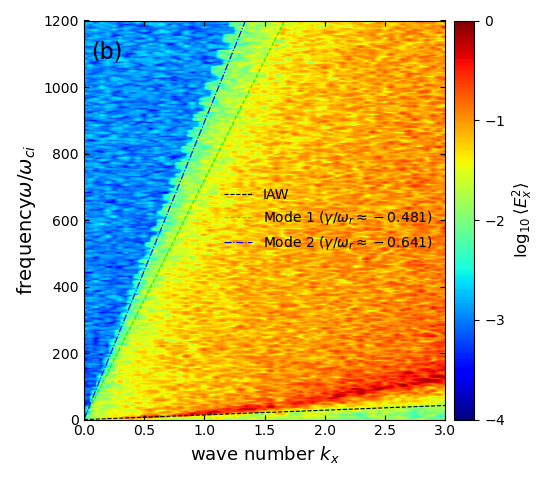}
    \includegraphics[width=0.45\linewidth]{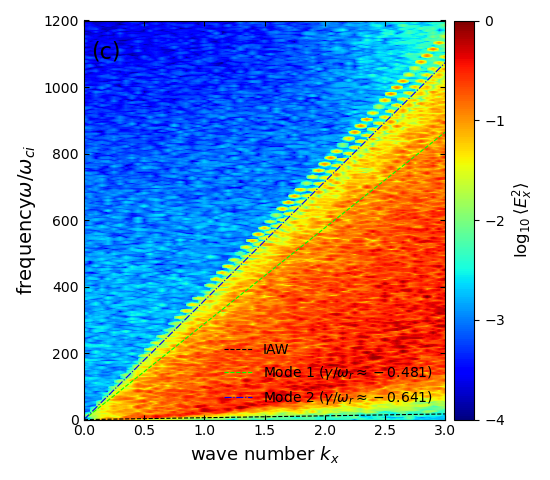}
    \includegraphics[width=0.45\linewidth]{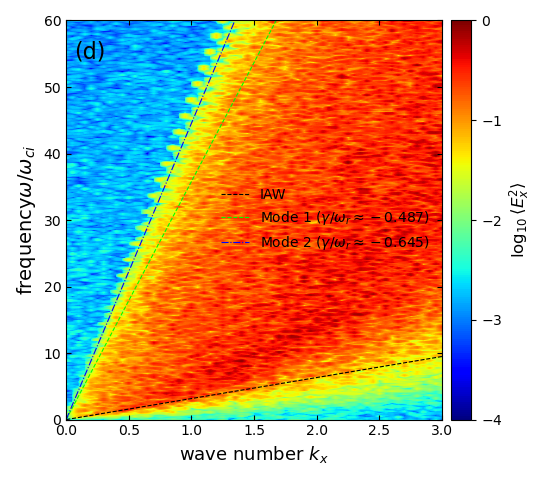}
    \caption{Numerical power spectra of the longitudinal electric field fluctuations from the DeFi-QN simulation. The power spectrum of $E_x$ ($E_\parallel$) is shown versus $k_x$ ($\kpar$). (a) Baseline power spectrum of $E_x$ fluctuations initialized from thermal noise, evaluated over an integration time $T=500$ ($\omega_{ci}T=2.5$). The physical parameters are $m_i/m_e = 200$, $T_i/T_e = 10^{-4}$, $v_{te} = 1$, and $\omega_{pe}=1$.  The simulation domain size is determined by $k_{\text{input}} = [0.05, 1]$ (where $L_{x,y} = 2\pi/k_{\text{input}}$), using a $256 \times 8$ grid with 400 particles per cell per species. 
    (b) Effect of increased spatial resolution using a $1024 \times 8$ grid, showing the shift of numerical deviations toward higher wavenumbers. (c) Spectrum for decreased electron thermal velocity ($v_{te}=0.4$), resulting in a downward frequency shift of high-order kinetic modes. (d) Impact of a reduced ion mass ($m_i=10$); the IAW branch shifts toward higher frequencies, consistent with the $\omega_{\rm IAW} \propto m_i^{-1/2}$ scaling, reduced ion mass leading to improved alignment with the lower boundary of the resonance zone.}
    \label{fig:GEMPICX_Ex_kx}
\end{figure}

The numerical spectrum of the longitudinal electric field fluctuations is shown in Fig. \ref{fig:GEMPICX_Ex_kx}(a). The fluctuations is initialized from thermal noise, evaluated over an integration time $T=500$ ($\omega_{ci}T=2.5$), $\Delta t=0.05$. The physical parameters are $m_i/m_e = 200$, $T_i/T_e = 10^{-4}$, $v_{te} = 1$, and $\omega_{pe}=1$ same as in Figs \ref{fig:ana_Ex_kx}.  The simulation domain size is determined by $k_{\text{input}} = [0.05, 1]$ (where $L_{x,y} = 2\pi/k_{\text{input}}$), using a $256 \times 8$ grid with 400 particles per cell per species. 
The quasi-neutral approximation effectively removes the high-frequency Langmuir branch, isolating the low-frequency kinetic response. In this regime, the thermal fluctuation spectrum is characterized by a ``triangular" resonance zone, bounded by the IAW branch and the higher-order damped modes.
Despite the presence of an elevated stochastic noise floor in the PIC simulations, the characteristic slopes of the thermal fluctuations (the upper boundary of the triangular region) show good agreement with the high-order roots derived from the Newton solver. This alignment demonstrates that the fundamental structure of the kinetic response is accurately captured, even in regions where discrete particle effects dominate the theoretical damping levels.
However, while the \texttt{GEMPICX} results show good agreement at low $k$, deviations appear along the lower boundary at intermediate to high wavenumbers. Specifically, within the triangular region, the numerical noise power at higher frequencies exceeds that associated with the IAW branch. 

To further investigate the origins of this noise, we perform a parametric study relative to the baseline case in Fig. \ref{fig:GEMPICX_Ex_kx}(a):
\begin{enumerate}
    \item Grid Resolution Effects: In Fig. \ref{fig:GEMPICX_Ex_kx}(b), the grid resolution is increased to $1024 \times 8$ (reducing the cell size $\Delta x$). As a result, the aforementioned deviations are shifted toward larger wavenumbers $k$, suggesting a resolution-dependent coupling numerical noise. While this case uses four times the total particle count (maintaining 400 particles per cell), additional sensitivity tests (not shown here) indicate that increasing the particle count alone reduces the absolute noise floor but does not shift the location of the deviations. This confirms that the observed spectral discrepancy is a numerical artifact related to the grid spacing rather than a lack of statistical particles.
    \item Electron Thermal Velocity: In Fig. \ref{fig:GEMPICX_Ex_kx}(c), the electron thermal velocity is set to $v_{te} = 0.4$. This leads to a decrease in the frequency of the IAW and the high-order modes, yet the upper boundary of the triangular region remains well-aligned with theoretical predictions. The location of the upper boundary depends on the electron temperature $T_e$, through the scaling $\omega_1 \propto v_{te}$.
    \item Ion Mass Ratio: Finally, Fig. \ref{fig:GEMPICX_Ex_kx}(d) examines the effect of a reduced ion mass ($m_i = 10$). In this case, the lower boundary of the fluctuation spectrum shows improved alignment with the IAW branch.
\end{enumerate}

\begin{figure}[htbp]
    \centering
    \includegraphics[width=0.5\linewidth]{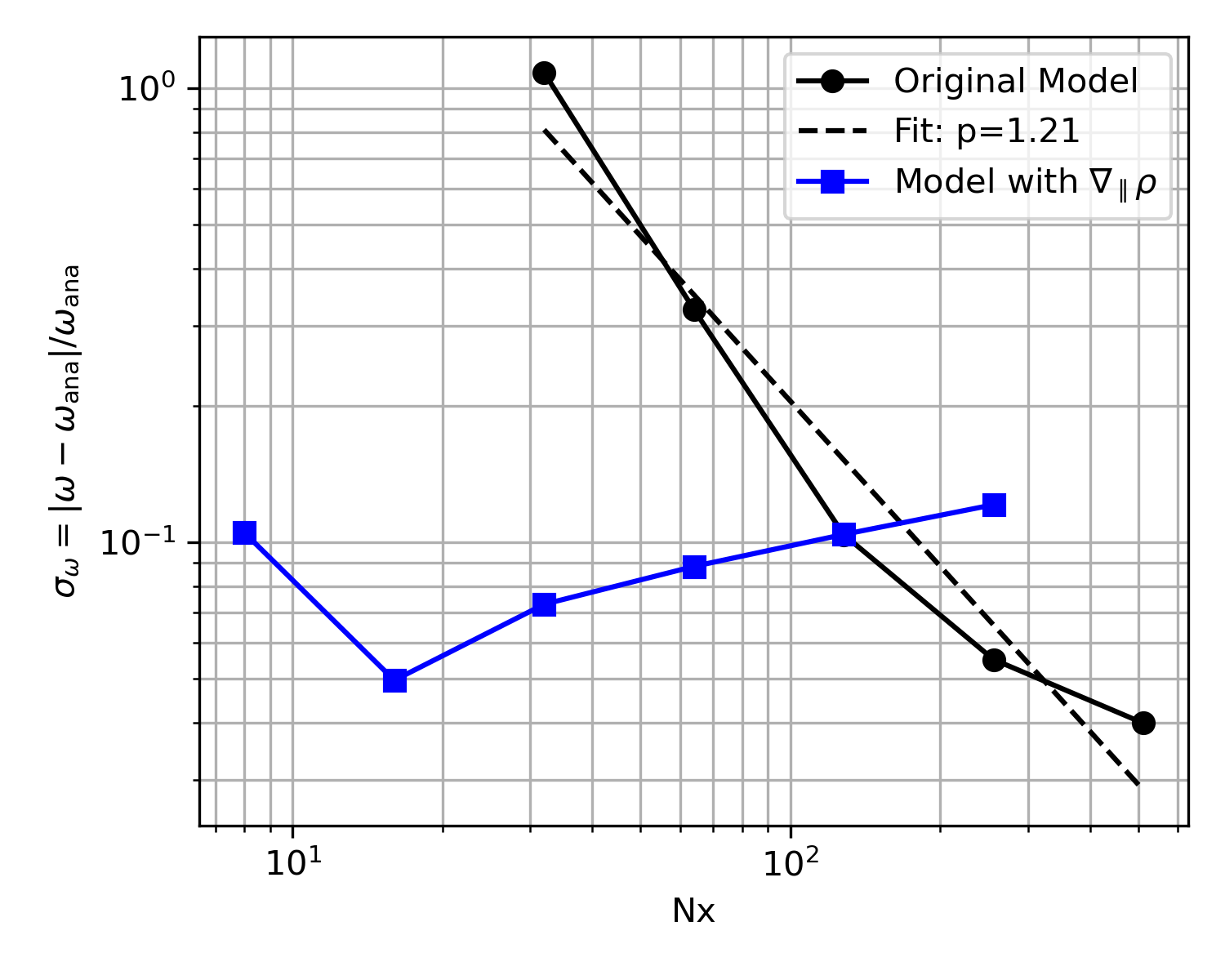}
    \caption{IAW frequency convergence vs $N_x$. }
    \label{fig:IAW_freq_error_VS_Nx}
\end{figure}

We adopt the parameter set $m_i/m_e = 200$, $T_i/T_e = 10^{-4}$, $v_{te} = 1$, and $k = 2\pi/10$, consistent with \cite{kormann2021energy}, where a one-dimensional simulation with 32 grid points and 128,000 particles per species was employed. In the present model, however, a higher spatial resolution is required due to increased numerical noise. Therefore, a convergence study of the frequency error was carried out by increasing the number of grid points in the parallel direction $N_x$ from 32 to 512, as illustrated in Fig. \ref{fig:IAW_freq_error_VS_Nx}. The simulations are initialized with a density perturbation. In the quasi-neutral model, charge neutrality requires $n_e = n_i$, and the initial condition is set as $n_e(t=0) = n_i(t=0) = 1 + 0.2 \cos(k_xx)$.

In our simulations, we use a time step $\Delta t = 0.05$, 400 particles per cell per species, and $N_y = 8$ grid points in the perpendicular direction. The analytical frequency is $\omega_{\mathrm{ana}}/\omega_{ci} = 8.8658$. With increasing resolution $N_x$ from 32 to 512,  the numerical frequency converges as 18.456, 11.750, 9.787, 9.355 and 9.220. The corresponding frequency error (black dots) is shown in Fig. \ref{fig:IAW_freq_error_VS_Nx}. We observe that the error initially decreases rapidly, followed by a much slower convergence at higher resolutions. A fitted convergence order of $p \approx 1.21$. The remaining error of approximately $4\%$ may be attributed to a combination of incomplete convergence, particle noise, finite time step effects, and uncertainties in frequency estimation.

This relatively low convergence rate poses a practical challenge for PIC  simulations. Since maintaining low numerical noise requires a sufficiently large number of particles per cell, and the time step  scales with the grid spacing ($\Delta t \propto \Delta x$), achieving convergence necessitates both increased particle numbers and reduced time steps, leading to significantly higher computational cost.

\begin{figure}[htbp]
    \centering
    \includegraphics[width=0.5\linewidth]{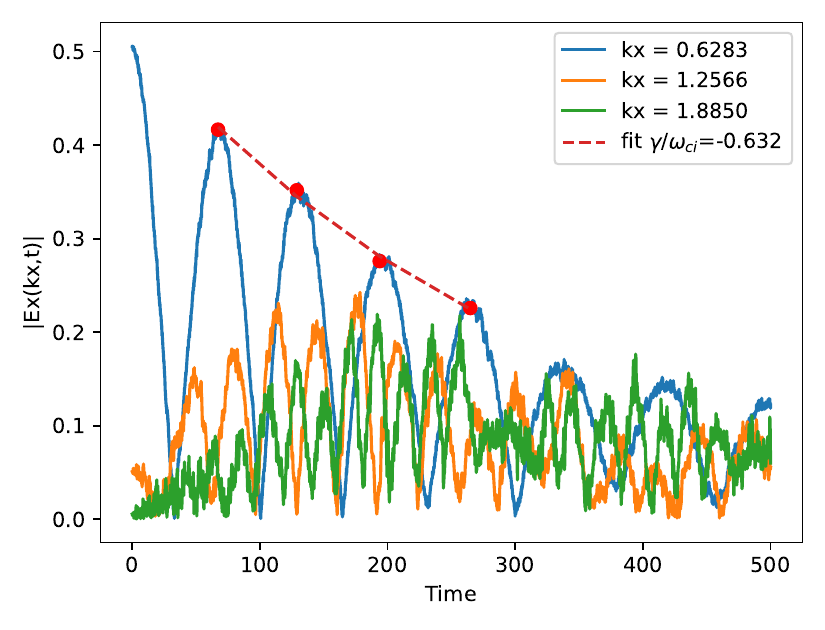}
    \caption{Time evolution of the first three $k_\parallel$ modes in the $N_x=512$ simulation.}
    \label{fig:IAW_k1k2k3}
\end{figure}

The time evolution of the first three $k_\parallel$ modes from $N_x=512$ simulation is shown in Fig. \ref{fig:IAW_k1k2k3}. The mode initialized by the density perturbation at $k = 2\pi/L$ exhibits a clear exponential decay in time, characteristic of IAW damping. A fit to the envelope gives a damping rate $\gamma/\omega_{ci} \approx -0.632$, which is larger in magnitude than the analytical value $\gamma_{\mathrm{ana}} = -0.39248$. In contrast, the higher-$k$ modes grow in time, indicating a transfer of energy from the primary mode to shorter wavelengths. This behavior is likely associated with numerical noise and mode coupling inherent to PIC simulations. As the primary mode damps, particle noise becomes relatively more important and can seed fluctuations at higher wavenumbers.
This suggests that the measured damping rate is affected by numerical noise, which can lead to an overestimation of damping rate.
\subsection{ Ion acoustic waves}
In the previous section, we demonstrated that noise fluctuations at higher frequencies exceed those associated with the IAW branch. Consequently, a large number of grid points and particles is required to obtain converged results in IAW simulations. This is requirement is undesirable, since one expects to be able to use fewer grids to resolve long wavelength modes.

Ion acoustic waves (IAWs) are challenging to resolve numerically due to the severe ``cancellation problem" encountered in the parallel Ohm's law (Eq.~\eqref{eq:curlcurlE}). This problem arises because numerical errors \cite{lu2025piecewise}  in the electron second-order moment term ($\frac{q_e}{m_e} (\nabla \cdot \mathbb{S}_e)\cdot \bex$ $\sim \nabla_\parallel p_{e \parallel}$) can be comparable in magnitude to the ion inertia contribution. As a result, the required cancellation between leading order terms ($\Epar$ from electron adiabatic response and $\nabla_\parallel p_{e \parallel}$) becomes inaccurate. This issue becomes more severe as the ion-to-electron mass ratio $m_i/m_e$ increases, as evidenced by Figs. \ref{fig:GEMPICX_Ex_kx}(a) compared with \ref{fig:GEMPICX_Ex_kx}(d).

As discussed in \cite{YChen2009VlasovDrift},  this issue can be mitigated by removing the adiabatic electron response, thereby enforcing exact cancellation of the leading-order perturbation terms.
We follow the same strategy by adding the term $-\mu_0\frac{  T_e}{m_e} \bex\cdot \nabla \rho$ to the right-hand side of Eq. \eqref{eq:curlcurlE}, where $\rho=\rho_i+\rho_e$ is the total charge density. 
For the IAWs with $\kperp=0$, purely electrostatic fields ($\bB=0$) and $\Eperp=0$, the parallel electric field equation becomes the only relevant field equation,
\begin{equation}
   \label{eq:Epar_IAW}
    (\frac{q_i^2}{m_i}n_i + \frac{q_e^2}{m_e}n_e ) \Epar =  \frac{q_i}{m_i} ( \nabla \cdot \mathbb{S}_i) \cdot \bex 
    + \frac{q_e}{m_e} (\nabla \cdot \mathbb{S}_e) \cdot \bex -\frac{   T_e}{m_e} \bex\cdot \nabla \rho 
\end{equation}
where  $(\nabla \cdot \mathbb{S}_i) \cdot \bex =\nabla_\parallel p_{i\parallel}$ and  $(\nabla \cdot \mathbb{S}_e) \cdot \bex =\nabla_\parallel p_{e \parallel}$ since $\Eperp=0$ and $\bB=0$. For a Maxwellian distribution, $\nabla_\parallel p_{s\parallel} = m_s v_{ts}^2 \nabla_\parallel n_s$. Consequently, the ratio of the second term to first is $v_{te}^2/v_{ti}^2=\frac{m_i}{m_e}\frac{T_e}{T_i}$, which is very large; therefore, the second term dominates in the IAW simulation.

In the ideal quasi-neutral limit, the total charge density $\rho = \rho_i + \rho_e$ vanishes identically. Consequently, this $\nabla\rho$ term acts as a numerical correction that subtracts residual charge errors arising from discrete particle noise and grid deposition. By introducing a term proportional to the gradient of the total charge density $\nabla \rho$, we effectively implement a numerical filter that suppresses high-frequency charge imbalances.

The error of IAW frequency is shown by the blue dot in Fig.~\ref{fig:IAW_freq_error_VS_Nx} with $N_x=8,\;16\;...\;256$. Notably, the inclusion of the $\nabla_\parallel\rho$ term significantly reduces frequency errors at lower grid resolutions, with an error less than $10\%$ on coarse meshes. However, for the corrected model (blue curve), the grid resolution is no longer the primary bottleneck for accuracy beyond $N_x \sim 16$. In this saturated regime, further improvements in precision would likely require optimizing other simulation parameters, such as reducing the time step size or increasing the particle number to minimize statistical noise.

\begin{figure}[htbp]
    \centering
    \includegraphics[width=0.8\linewidth]{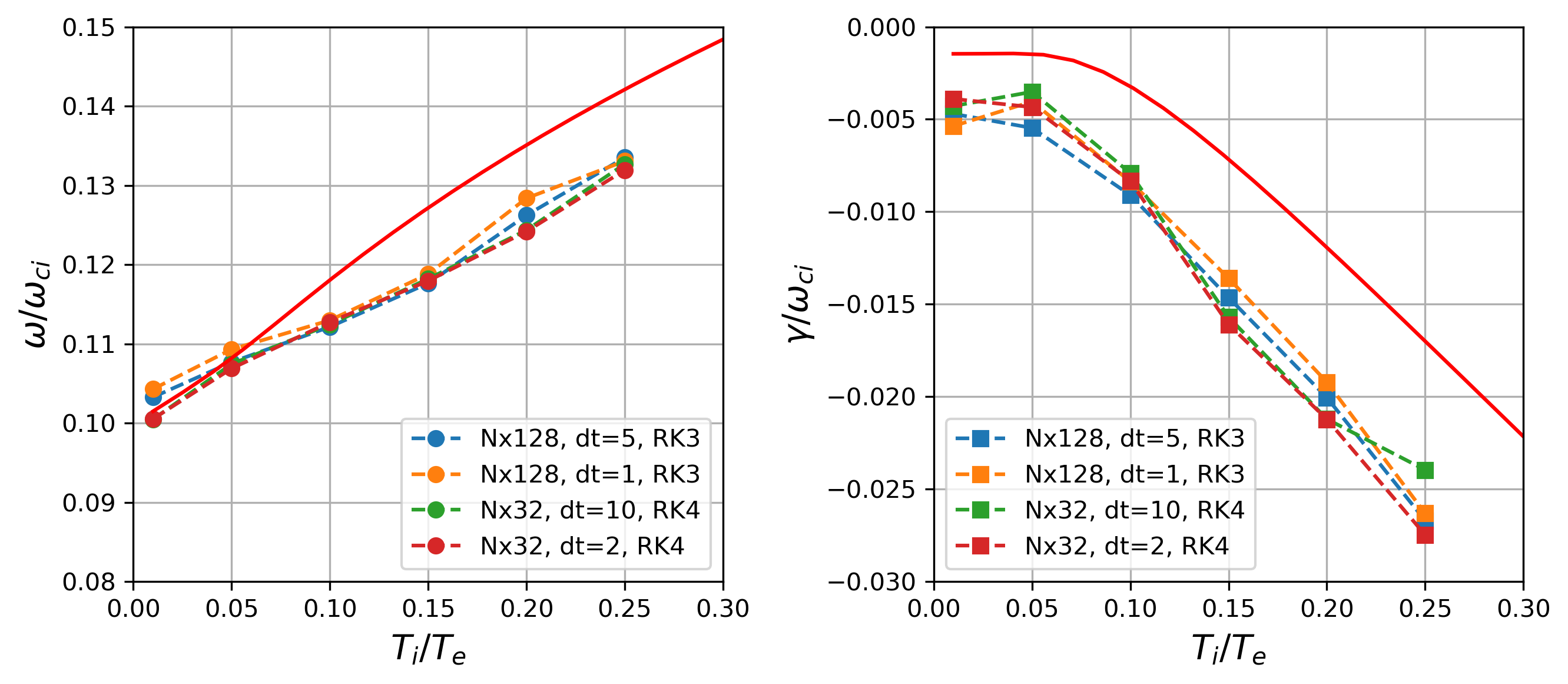}
    \caption{Frequency and damping rate of the ion acoustic wave for a realistic mass ratio $m_i/m_e=1836$. The red solid line represents the analytical solution of Eq. \eqref{eq:QN_Dzz}.}
    \label{fig:IAW_realistic}
\end{figure}

The inclusion of the $\nabla_\parallel\rho$ term facilitates efficient IAW simulations at large mass ratios. Here, we provide benchmark results for a proton-electron plasma. The frequency and damping rate of IAW with $v_{te}=0.1$, $\kpar=0.1/\rho_{\rm sound}$ are shown in Fig.~\ref{fig:IAW_realistic}. We set $k_{\rm input}=[0.1/\rho_{\rm sound}, 0.01]$. The domain is discretized using a $N_x\times 8$ cell with 4000 particles per cell per species. The simulation is initialized with a density perturbation of the form $n_e=n_i=1+0.2 \cos(k_xx)$.  Other parameters grid number $N_x$, time step $\Delta t$, and LSRK method are listed on the figure.
As shown in Fig.~\ref{fig:IAW_realistic}, both the frequency and damping rate follow the analytical trends. Quantitatively, the frequency error remains below $10\%$, while the damping rate exhibits a noticeable offset. The enhanced damping may be attributed to contributions from high-$k$ modes, as discussed for the reduced case in Fig.~\ref{fig:IAW_k1k2k3}. While the frequency results are converged at a coarse resolution of $N_x=32$ and $\Delta t=10$, the damping rate requires a smaller time step for convergence. Furthermore, the discrepancy between numerical and analytical results increases with $T_i/T_e$; this is likely due to the higher damping rates, which lead to a rapid decay of the signal. As a result, the signal-to-noise ratio decreases, causing the measured damping rate to deviate.

\section{Conclusion}\label{sec:conclusion}
In this work, we present a consistent derivation of the DeFi-QN model starting from the drift-kinetic electron and fully kinetic ion Vlasov equations. By enforcing quasi-neutrality, the model eliminates high-frequency light waves and Langmuir oscillations, providing an efficient electromagnetic framework specifically tailored for low-frequency plasma dynamics. Analysis of the cold plasma dispersion relations shows that the DeFi-QN model retains a high-frequency branch with a quadratic dependence on $k$, which connects the compressional Alfv\'en wave to whistler-wave physics in the long-wavelength limit. While physically well-grounded at small $k$, this branch introduces numerical stiffness and imposes a severe time step constraint. The validity regime of the model is discussed in detail in terms of the drift-kinetic ordering and quasi-neutral assumptions. We use explicit low-storage Runge-Kutta schemes for the time discretization and a dual-grid approach for the spatial discretization. To address the numerical stiffness inherent in the fast whistler branch, we propose a novel implicit-explicit (IMEX) splitting scheme for advancing the magnetic field. The stiff term is shown to be intrinsically linked to Hall physics and the underlying whistler dynamics, providing a clear physical interpretation of the numerical constraint and its algorithmic solution. 
The formulation was benchmarked against analytical predictions for a wide range of modes, including ion Bernstein waves, as well as shear and compressional Alfv\'en waves. The results confirm that the model accurately captures relevant low-frequency electromagnetic phenomena within its regime of validity. Notably, simulations in the limit of small electron skin depth demonstrate that the $\mathbf{E}$-$\mathbf{B}$ formulation effectively mitigates the cancellation problems typically encountered in shear Alfv\'en wave simulations using traditional potential-based models. Furthermore, our investigation of longitudinal fluctuations and ion acoustic waves (IAW) highlights the model's reliability in the electrostatic limit, while also identifying performance boundaries related to the parallel Ohm's law.  

The model is well suited for studying interactions between energetic particles and Alfv\'en waves, ion cyclotron range of frequencies phenomena, non-linear wave-particle resonances, and multi-ion plasmas \cite{chen2016RMP, gorelenkov1995ACI_ICE, chen2010linear, meng2018resonance}. Although the current full-$f$ PIC implementation is subject to the inherent statistical noise associated with large marker counts, it enables a fully self-consistent treatment of background variations. The DeFi-QN framework establishes a solid foundation for future extensions. Future work will focus on implementing a control variate (direct-$\delta f$) method to enhance the signal-to-noise ratio, as well as exploring mixed $\delta f$-full-$f$ approaches \cite{lu2023full}. In addition, collisional effects will be incorporated to extend the model's applicability to more collisional plasma regimes such as edge plasmas \cite{sonnendrucker2015split}.

\ack{ Guo Meng acknowledges valuable discussions with Dr. Yingzhe Li and Dr. Omar Maj. The authors would like to thank the GEMPICX team members for helpful discussions, code development, and technical assistance during this work.
The EUROfusion project TSVV-G is acknowledged.  
This work has been carried out within the framework of the EUROfusion Consortium, funded by the European Union via the Euratom Research and Training Programme (Grant Agreement No 101052200-EUROfusion). Views and opinions expressed are, however, those of the author(s) only and do not necessarily reflect those of the European Union or the European Commission. Neither the European Union nor the European Commission can be held responsible for them.}

\bibliographystyle{iopart-num}
\bibliography{gempic-qn-dk}

@article{kraus2017gempic,
	author = {Kraus, Michael and Kormann, Katharina and Morrison, Philip J and Sonnendr{\"u}cker, Eric},
	journal = {Journal of Plasma Physics},
	number = {4},
	pages = {905830401},
	publisher = {Cambridge University Press},
	title = {{GEMPIC: geometric electromagnetic particle-in-cell methods}},
	volume = {83},
	year = {2017}}

@article{Kormann2024A-Dual-Grid,
	author = {Kormann, Katharina and Sonnendr\"{u}cker, Eric},
	journal = {SIAM Journal on Scientific Computing},
	number = {5},
	pages = {B621-B646},
	title = {A Dual Grid Geometric Electromagnetic Particle in Cell Method},
	volume = {46},
	year = {2024}}

@article{campos2022variational,
  title={Variational framework for structure-preserving electromagnetic particle-in-cell methods},
  author={Campos Pinto, Martin and Kormann, Katharina and Sonnendr{\"u}cker, Eric},
  journal={Journal of Scientific Computing},
  volume={91},
  number={2},
  pages={46},
  year={2022},
  publisher={Springer}
}

@software{gempicx,
  title = {GEMPICX: Geometric Electro-Magnetic Particle-In-Cell for eXascale},
  author = {GEMPICX-Team},
  year = {2026},
  url = {https://github.com/NMPPMaxPlanck/GEMPICX}
}

@article{sonnendrucker2015split,
  title={A split control variate scheme for PIC simulations with collisions},
  author={Sonnendr{\"u}cker, Eric and Wacher, Abigail and Hatzky, Roman and Kleiber, Ralf},
  journal={Journal of Computational Physics},
  volume={295},
  pages={402--419},
  year={2015},
  publisher={Elsevier}
}

@article{Meng2025,
author = {Meng, Guo and Kormann, Katharina and Poulsen, Emil and Sonnendr{\"u}cker, Eric},
year = {2025},
month = {04},
pages = {055007},
title = {A geometric Particle-In-Cell discretization of the drift-kinetic and fully kinetic Vlasov--Maxwell equations},
volume = {67},
journal = {Plasma Physics and Controlled Fusion},
doi = {10.1088/1361-6587/adc832}
}

@book{stix1992waves,
  title={Waves in plasmas},
  author={Stix, Thomas H},
  year={1992},
  publisher={Springer Science \& Business Media}
}

@book{chen1987waves,
  title={Waves and instabilities in plasmas},
  author={Chen, Liu},
  volume={12},
  year={1987},
  publisher={World scientific}
}

@article{Degond2017AP-review,
	author = {Pierre Degond and Fabrice Deluzet},
	journal = {Journal of Computational Physics},
	pages = {429-457},
	title = {Asymptotic-Preserving methods and multiscale models for plasma physics},
	volume = {336},
	year = {2017}}

@article{littlejohn1983variational,
  title={Variational principles of guiding centre motion},
  author={Littlejohn, Robert G},
  journal={Journal of Plasma Physics},
  volume={29},
  number={1},
  pages={111--125},
  year={1983},
  publisher={Cambridge University Press}
}

@article{brizard2007foundations,
	author = {Brizard, A. J. and Hahm, T. S.},
	journal = {Rev. Mod. Phys.},
	month = {Apr-Jun},
	number = {2},
	pages = {421--468},
	title = {Foundations of nonlinear gyrokinetic theory},
	volume = {79},
	year = {2007}}

@article{garbet2010gkturb,
  title={Gyrokinetic simulations of turbulent transport},
  author={Garbet, Xavier and Idomura, Yasuhiro and Villard, Laurent and Watanabe, TH},
  journal={Nuclear Fusion},
  volume={50},
  number={4},
  pages={043002},
  year={2010}
}

@article{nishant2026,
  title={Geometric numerical discretization of electromagnetic quasineutral models},
  author={Narechania, N. and Poulsen, E. and Sonnendr\"ucker, E.},
  journal={Journal of Plasma Physics},
  volume={submitted},
  year={2026},
  publisher={Cambridge University Press}
}

@article{nishant_cpc,
  title={Geometric numerical discretization of a quasi-neutral hybrid
model of drift-kinetic electrons and fully kinetic ions},
  author={Narechania, N. and Meng, G. and Poulsen, E. and Sonnendr\"ucker, E.},
  journal={Journal of Computational Physics},
  volume={submitted},
  year={2026}
}

@article{bao2014LHW,
  title={Particle simulation of lower hybrid wave propagation in fusion plasmas},
  author={Bao, J and Lin, Z and Kuley, A and Lu, ZX},
  journal={Plasma Physics and Controlled Fusion},
  volume={56},
  number={9},
  pages={095020},
  year={2014},
  publisher={IOP Publishing}
}

@article{kuley2013verification,
  title={Verification of particle simulation of radio frequency waves in fusion plasmas},
  author={Kuley, Animesh and Wang, ZX and Lin, Z and Wessel, F},
  journal={Physics of Plasmas},
  volume={20},
  number={10},
  year={2013},
  publisher={AIP Publishing}
}

@article{bao2018conservative,
  title={A conservative scheme for electromagnetic simulation of magnetized plasmas with kinetic electrons},
  author={Bao, J and Lin, Z and Lu, ZX},
  journal={Physics of Plasmas},
  volume={25},
  number={2},
  year={2018},
  publisher={AIP Publishing}
}

@article{lin2005gefi,
  title={Three-dimensional global hybrid simulation of dayside dynamics associated with the quasi-parallel bow shock},
  author={Lin, Yu and Wang, XY},
  journal={Journal of Geophysical Research: Space Physics},
  volume={110},
  number={A12},
  year={2005},
  publisher={Wiley Online Library}
}

@article{chen2019gefiEB,
  title={A new particle simulation scheme using electromagnetic fields},
  author={Chen, L and Lin, Y and Wang, XY and Bao, J},
  journal={Plasma Physics and Controlled Fusion},
  volume={61},
  number={3},
  pages={035004},
  year={2019},
  publisher={IOP Publishing}
}

@article{rosen2022EB,
  title={An E and B gyrokinetic simulation model for kinetic Alfv{\'e}n waves in tokamak plasmas},
  author={Rosen, MH and Lu, ZX and Hoelzl, Matthias},
  journal={Physics of Plasmas},
  volume={29},
  number={2},
  year={2022},
  publisher={AIP Publishing}
}

@article{YChen2009VlasovDrift,
  title={Particle-in-cell simulation with Vlasov ions and drift kinetic electrons},
  author={Chen, Yang and Parker, Scott E},
  journal={Physics of Plasmas},
  volume={16},
  number={5},
  year={2009},
  publisher={AIP Publishing}
}

@article{lu2021implicit,
  title={The development of an implicit full f method for electromagnetic particle simulations of Alfv{\'e}n waves and energetic particle physics},
  author={Lu, ZX and Meng, G and Hoelzl, Matthias and Lauber, Ph},
  journal={Journal of Computational Physics},
  volume={440},
  pages={110384},
  year={2021},
  publisher={Elsevier}
}

@article{lu2023full,
  title={Full f and $$\backslash$delta f $ gyrokinetic particle simulations of Alfv{\'e}n waves and energetic particle physics},
  author={Lu, Zhixin and Meng, Guo and Hatzky, Roman and Hoelzl, Matthias and Lauber, Philipp},
  journal={Plasma Physics and Controlled Fusion},
  volume={65},
  number={3},
  pages={034004},
  year={2023},
  publisher={IOP Publishing}
}

@article{lu2025generalized,
  title={Generalized mixed variable-pullback scheme with ideal/non-ideal Ohm's law and pure symplectic scheme for electromagnetic gyrokinetic simulations},
  author={Lu, Zhixin and Meng, Guo and Hatzky, Roman and Sonnendr{\"u}cker, Eric and Mishchenko, Alexey and Hoelzl, Matthias},
  journal={Physics of Plasmas},
  volume={32},
  number={12},
  year={2025},
  publisher={AIP Publishing}
}

@article{lu2025piecewise,
  title={Piecewise field-aligned finite element method for multi-mode nonlinear particle simulations in tokamak plasmas},
  author={Lu, Zhixin and Meng, Guo and Sonnendr{\"u}cker, Eric and Hatzky, Roman and Mishchenko, Alexey and Zonca, Fulvio and Lauber, Philipp and Hoelzl, Matthias},
  journal={Journal of Plasma Physics},
  volume={91},
  number={2},
  pages={E48},
  year={2025},
  publisher={Cambridge University Press}
}

@article{araneda2012interactions,
  title={Interactions of Alfv{\'e}n-cyclotron waves with ions in the solar wind},
  author={Araneda, JA and Astudillo, H and Marsch, E},
  journal={Space science reviews},
  volume={172},
  number={1},
  pages={361--372},
  year={2012},
  publisher={Springer}
}

@book{sitenko1967,
  author    = {A. G. Sitenko},
  title     = {Electromagnetic Fluctuations in Plasma},
  publisher = {Academic Press},
  address   = {New York},
  year      = {1967},
  note      = {See p.~28, Eq.~(2.41)}
}

@article{mishchenko2014pullback,
  title={Pullback transformation in gyrokinetic electromagnetic simulations},
  author={Mishchenko, Alexey and K{\"o}nies, Axel and Kleiber, Ralf and Cole, Michael},
  journal={Physics of Plasmas},
  volume={21},
  number={9},
  year={2014},
  publisher={AIP Publishing}
}

@book{cummings1994gyrokinetic,
  title={Gyrokinetic simulation of finite-beta and self-generated sheared-flow effects on pressure-gradient-driven instabilities},
  author={Cummings, Julian Clark},
  year={1994},
  publisher={Princeton University}
}

@article{chen2011energy,
  title={An energy-and charge-conserving, implicit, electrostatic particle-in-cell algorithm},
  author={Chen, Guangye and Chac{\'o}n, Luis and Barnes, Daniel C},
  journal={Journal of Computational Physics},
  volume={230},
  number={18},
  pages={7018--7036},
  year={2011},
  publisher={Elsevier}
}

@article{kormann2021energy,
  title={Energy-conserving time propagation for a structure-preserving particle-in-cell Vlasov--Maxwell solver},
  author={Kormann, Katharina and Sonnendr{\"u}cker, Eric},
  journal={Journal of Computational Physics},
  volume={425},
  pages={109890},
  year={2021},
  publisher={Elsevier}
}

@article{gorelenkov1995ACI_ICE,
  title={Alfv{\'e}n cyclotron instability and ion cyclotron emission},
  author={Gorelenkov, NN and Cheng, CZ},
  journal={Nuclear fusion},
  volume={35},
  number={12},
  pages={1743--1752},
  year={1995}
}

@article{chen2016RMP,
  title={Physics of Alfv{\'e}n waves and energetic particles in burning plasmas},
  author={Chen, Liu and Zonca, Fulvio},
  journal={Reviews of Modern Physics},
  volume={88},
  number={1},
  pages={015008},
  year={2016},
  publisher={APS}
}

@article{meng2018resonance,
  title={Resonance frequency broadening of wave-particle interaction in tokamaks due to Alfv{\'e}nic eigenmode},
  author={Meng, Guo and Gorelenkov, Nikolai N and Duarte, Vinicius N and Berk, Herb L and White, Roscoe B and Wang, X. G},
  journal={Nuclear Fusion},
  volume={58},
  number={8},
  pages={082017},
  year={2018},
  publisher={IOP Publishing}
}

@article{chen2010linear,
  title={Linear gyrokinetic simulation of high-n toroidal Alfv{\'e}n eigenmodes in a burning plasma},
  author={Chen, Yang and Parker, Scott E and Lang, J and Fu, G-Y},
  journal={Physics of Plasmas},
  volume={17},
  number={10},
  year={2010},
  publisher={AIP Publishing}
}

\end{document}